\documentclass[aip,jcp,reprint,preprintnumbers,amsmath,amssymb,floatfix,nofootinbib]{revtex4-1}

\usepackage{graphicx}

\usepackage{amsmath}
\usepackage{mathtools}
\usepackage{amssymb}
\usepackage{amsfonts}  
\usepackage{rotating}
\usepackage{multirow}
\usepackage{dsfont}
\newcommand{\braket}[2]{\langle #1|#2\rangle}

\newcommand{\rvec}{\mathbf{R}}

\newcommand{\rv}{\mathbf{r}}
\newcommand{\pv}{\mathbf{p}}
\newcommand{\vv}{\mathbf{v}}

\newcommand{\xvec}{\mathbf{x}}

\newcommand{\xdot}{\dot{x}}

\newcommand\numberthis{\addtocounter{equation}{1}\tag{\theequation}}

\newcommand{\allq}{\{\rvec^{(\alpha)}\} }

\usepackage{color}

\begin{document}
\edef\marginnotetextwidth{\the\textwidth}

\title{Kinetically Constrained Ring-Polymer Molecular Dynamics for Non-adiabatic Chemical Reactions} 

\author{Artur R. Menzeleev}
\author{Franziska Bell}
\author{Thomas F. Miller III}
\email{tfm@caltech.edu}

\affiliation{
Division of Chemistry and Chemical Engineering, California Institute of Technology, Pasadena, CA 91125, USA
}

\date{\today}
\begin{abstract}
We extend  ring-polymer molecular dynamics (RPMD)  to allow for the  direct simulation of general, electronically non-adiabatic chemical processes.
The kinetically constrained  (KC) RPMD method uses the imaginary-time path-integral representation in the set of nuclear coordinates and electronic states to provide 
continuous equations of motion that describe the quantized, electronically non-adiabatic 
dynamics of the system.
 KC-RPMD  preserves 
the favorable  properties of the 
usual RPMD formulation in the position representation, including rigorous detailed balance, time-reversal symmetry, and  invariance of reaction rate calculations to the choice of  dividing surface.
However, the new method overcomes  significant shortcomings of  position-representation RPMD  
by enabling the description of non-adiabatic transitions between states associated with general, many-electron wavefunctions and by accurately describing deep-tunneling processes across asymmetric barriers.
We demonstrate that  KC-RPMD  yields excellent numerical results for a range of model systems, including a simple avoided-crossing reaction and condensed-phase electron-transfer reactions across multiple regimes for the electronic coupling and thermodynamic driving force.
\end{abstract}

\maketitle

\section{Introduction}
A central challenge in chemical dynamics is the accurate and robust description of non-adiabatic processes in the condensed phase.
Important target applications include charge-transfer and energy-transfer processes that are fundamental to biological and inorganic catalysis. 
A variety of simulation methods have been developed to address this challenge, including those based on mean-field,\cite{Ehr27,Mey79,Micha83,Tully98, Truhlar00} surface hopping,\cite{Tully71,Tully90,Kuntz91} and semiclassical dynamics\cite{LSCIVRone,LSCIVRtwo,Cot13,Frank13} approaches.
In the current study, we provide a novel extension of the ring-polymer molecular dynamics (RPMD) method that is well suited to addressing   electronically non-adiabatic dynamics and nuclear quantization for chemical reactions in large systems.

RPMD is an approximate quantum dynamics method\cite{Cra04, tfm2013} that is based on the path-integral formalism of statistical mechanics.\cite{Feynman} It provides an isomorphic classical model for the real-time evolution of a quantum mechanical system. 
RPMD yields real-time molecular dynamics trajectories that preserve the exact quantum Boltzmann distribution and exhibit time-reversal symmetry, thus enabling the method to be readily used in combination with classical rare-event sampling methods and for the direct simulation of quantum-mechanical processes in systems involving thousands of atoms. 
Numerous applications of the RPMD method have been reported to date,\cite{tfm2013} including the study of chemical reactions in the gas phase,\cite{Col09,deT12,Sul11,All13} in solution,\cite{Cra05,Cra05b,Col08,Men11,Kre13} and in enzymes;\cite{Boe11}
the simulation of diffusive processes in liquids,\cite{tfm05rpmdb,tfm05rpmdc,tfm08,Hab09,Hab09b,Mar08,Men10} glasses,\cite{Mar11,Mar12} solids,\cite{Mar08} and on surfaces;\cite{Cal10,Sul12} and the calculation of neutron diffraction patterns \cite{Cra06} and absorption spectra.\cite{Hab08, Shi08}

We have recently employed the RPMD method to investigate condensed-phase electron transfer (ET)\cite{Men11} and proton-coupled electron transfer (PCET)\cite{Kre13} reaction dynamics.  
This work utilized the usual path-integral formulation in the position representation,\cite{Cha81,Feynman,Par84,Rae84} 
 such that the transferring electron is treated as a distinguishable particle. 
Although this approach allows for the robust description of condensed-phase charge transfer, it is clearly limited to non-adiabatic processes that  can be realistically described using a one-electron pseudopotential, rather than general, many-electron wavefunctions.\cite{Men11,Kre13}
Recent efforts have been made to extend RPMD to more general non-adiabatic chemistries, 
such as combining the path-integral methods with fewest-switches surface hopping \cite{Shu12} or  approaches \cite{Ana10,Ric13,Ana13} based on the  Meyer-Miller-Stock-Thoss Hamiltonian.\cite{Mey79,Sto97} However, the development of electronic-state-representation (or simply ``state-representation") RPMD methods that provide  accuracy and scalability while strictly preserving detailed balance 
remains an ongoing challenge.

In this work, we extend  RPMD  to allow for the description of non-adiabatic, multi-electron processes in large systems. 
The new  kinetically constrained (KC) RPMD method
employs a coarse-graining procedure that reduces discrete electronic-state variables to a single continuous coordinate, as well as a ``kinetic constraint" modification of the equilibrium distribution to
address known failures of path-integral-based estimates for tunneling rates. 
This kinetically constrained distribution is rigorously preserved using continuous equations of motion, yielding a real-time model for the non-adiabatic dynamics that retains 
all the useful features of the conventional position-representation  RPMD method, such as detailed balance, time-reversal symmetry, and invariance of reaction rate calculations to the choice of  dividing surface. We demonstrate that the method yields excellent numerical results for a range of model systems, including a simple avoided-crossing reaction and condensed-phase ET reactions across multiple regimes for the electronic coupling and thermodynamic driving force.

\section{Theory}
\label{sec:theory}

\subsection{Path-integral discretization in a two-level system}
We begin by reviewing imaginary-time path integration for 
a general, two-level system in the diabatic representation. 
Consider a Hamiltonian operator of the form $\hat{H} = \hat{T} + \hat{V}$, where
\begin{equation}
\hat{T} = \sum_{j=1}^{d} \frac{p^2_j}{2m_j}
\end{equation}
describes the  kinetic energy for a system of $d$ nuclear degrees of freedom and 
\begin{equation}
\label{eq:potmat}
\hat{V}(\rvec) = 
\begin{pmatrix}
V_{0}(\rvec) &  K(\rvec) \\
K(\rvec) & V_{1}(\rvec)  \end{pmatrix}
\end{equation}
is the potential energy in the diabatic representation as a function of the 
nuclear coordinates, $\rvec$.

The canonical partition function for the two-level system is
\begin{align*}
\label{eq:pi_z}
Z &=  \textrm{Tr} [e^{-\beta\hat{H}}]\\
&=  \int d\rvec \sum_{i=0,1}  \langle \rvec, i | e^{-\beta \hat{H}} | \rvec, i \rangle. \numberthis
\end{align*}
By resolving the identity in the product space of the electronic and nuclear coordinates, we discretize the trace into the ring-polymer representation with $n$ beads,
\begin{equation}
Z \!=\!\!  \int\! d \allq \! \sum_{\{i^{(\alpha)}\}}  \prod_{\alpha=1}^n \langle \rvec^{(\alpha)}, i^{(\alpha)} | e^{-\beta_n \hat{H}} | \rvec^{(\alpha+1)}, i^{(\alpha+1)} \rangle,
\end{equation}
where $\beta_n=\beta/n$ and 
$\left(\rvec^{(\alpha)},i^{(\alpha)}\right)$
indicates the nuclear position and electronic state of the 
$\alpha^{\textrm{th}}$
ring-polymer bead, such that 
$\left(\rvec^{(n+1)},i^{(n+1)}\right) = \left(\rvec^{(1)},i^{(1)}\right)$. 
Finally, employing the short-time approximations 
\begin{equation}
 \langle \rvec, i | e^{-\beta_n \hat{H}} | \rvec', i' \rangle \approx 
 \langle \rvec | e^{-\beta_n \hat{T}} | \rvec' \rangle  
  \langle i | e^{-\beta_n \hat{V}\left(\rvec\right)} | i' \rangle
\end{equation}
 and 
\begin{equation}
\label{eq:sta1}
\langle i | e^{-\beta_n \hat{V}(\rvec)} | i' \rangle \approx 
[\mathbf{M}(\rvec)]_{i,i'},
\end{equation}
where \cite{GreenBook}  
\begin{equation}
\label{eq:sta2}
\mathbf{M}(\rvec)\! = \!\!
\begin{pmatrix}
e^{-\beta_n V_{0}(\rvec)} & - \beta_n K(\rvec) e^{-\beta_n V_{0}(\rvec)}  \\
-\beta_n K(\rvec) e^{-\beta_n V_{1}(\rvec)}  & e^{-\beta_n V_{1}(\rvec)}  \end{pmatrix},
\end{equation}
we obtain the familiar result,
\begin{equation}
\label{eq:standardPF}
Z_n =\int\!\! d\{\rvec^{(\alpha)}\}\!\!  \sum_{\{i^{(\alpha)}\}}
\rho^{\textrm{RP}}_n(\{\rvec^{(\alpha)}\},\{i^{(\alpha)}\}),
\end{equation}
such that $Z\! =\! \lim_{n\rightarrow \infty} Z_n$. 
The ring-polymer distribution in Eq. \ref{eq:standardPF} is given by
\begin{eqnarray}
\label{eq:standardPIdist}
&&\rho^{\textrm{RP}}_n(\{\rvec^{(\alpha)}\},\!\{i^{(\alpha)}\})= \nonumber \\
&&\qquad\qquad\Omega e^{-\beta U_{\mathrm{int}} (\allq)} 
\prod_{\alpha=1}^n \!M_{i^{(\alpha)},i^{(\alpha+1)}}(\rvec^{(\alpha)}).
\end{eqnarray}
Here, we have introduced the notation $\Omega~=~\prod_{j=1}^{d} {\left(\frac{n m_j}{2\pi\hbar^2\beta}\right)^{n/2}}$ and $[\mathbf{M}(\rvec)]_{i,i'}~=~M_{i,i'}(\rvec)$,
as well as the internal ring-polymer potential 
\begin{equation}
U_{\mathrm{int}} (\allq)  =  \frac{1}{2n}\sum_{\alpha=1}^n \sum_{j=1}^{d}  
m_j\omega_{n}^2\left(R^{(\alpha)}_j-R^{(\alpha+1)}_j\right)^2,
\end{equation}
 where $\omega_{n}=(\beta_n\hbar)^{-1}$.

\subsection{Mean-field (MF) non-adiabatic RPMD}
\label{sec:eomMF}

Equation \ref{eq:standardPF} can be rewritten in the form of a classical configuration integral,
\begin{equation}
\label{eq:pf_MF}
Z_n=\int\!\! d\{\rvec^{(\alpha)}\}\   \rho_n^{\mathrm{MF}}(\allq),
\end{equation}
where  $\rho_n^{\mathrm{MF}}(\allq)$ is a quantized equilibrium distribution that depends only on the ring-polymer nuclear coordinates,
\begin{equation}
\label{eq:rho_MF}
 \rho_n^{\mathrm{MF}}(\allq)=
 \Omega  
 e^{-\beta V_{\mathrm{eff}}^{\mathrm{MF}}(\allq)},
\end{equation}
and 
\begin{eqnarray}
\label{eq:mf_veff}
V_{\mathrm{eff}}^{\mathrm{MF}}(\allq&&)=U_{\mathrm{int}} (\allq) \\ 
&&-\frac{1}{\beta}\ln \left[ \sum _{\{i^{(\alpha)}\}} 
 \prod_{\alpha=1}^n M_{i^{(\alpha)},i^{(\alpha+1)}}(\rvec^{(\alpha)}) \right]. \nonumber
\end{eqnarray}
Here, $V_{\mathrm{eff}}^{\mathrm{MF}}(\allq)$ is an effective potential for the ring-polymer nuclear coordinates in which all fluctuations over the electronic state variables are thermally averaged; in this sense, it provides a  mean-field (MF) description of the electronic degrees of freedom.

As is familiar from applications of path-integral statistical mechanics,\cite{Par84,Rae84}
the quantized equilibrium distribution 
can be sampled by running appropriately thermostatted classical molecular dynamics trajectories on the 
effective ring-polymer potential. 
Specifically, the classical equations of motion that sample $\rho_n^{\mathrm{MF}}(\allq)$ are 
\begin{equation}
\label{eq:mf_eomsA}
\dot{v}_j^{(\alpha)}= -\frac{1}{\tilde{m}_j} \frac{\partial}{\partial R_j^{(\alpha)}} V^{\mathrm{MF}}_{\mathrm{eff }}(\allq). 
\end{equation}
We use a notation for the masses in Eq. \ref{eq:mf_eomsA} 
that emphasizes that they need not 
correspond to the physical masses of the system; any positive values for these masses will yield trajectories 
that correctly sample the path-integral distribution.  However, to employ these trajectories 
as a model for the real-time dynamics of the system, it is sensible, as in previous implementations of RPMD,\cite{tfm2013} to utilize masses for the nuclear degrees of freedom that 
correspond to the physical masses of the system (i.e., $\tilde{m}_j=m_j/n$). 
This choice 
is sufficient to fully specify the MF 
version of non-adiabatic  RPMD dynamics for two-level systems, 
\begin{equation}
\label{eq:mf_eoms}
\begin{split}
\dot{v}_j^{(\alpha)}= -\frac{n}{m_j} \frac{\partial}{\partial R_j^{(\alpha)}} V^{\mathrm{MF}}_{\mathrm{eff }}(\allq).
\end{split}
\end{equation}

MF non-adiabatic RPMD, 
described in Eq. \ref{eq:mf_eoms},  
has the appealing feature that it involves simple, continuous equations of motion that rigorously preserve the exact quantum Boltzmann distribution.\cite{NandiniNote} 
However, as we will illustrate with later results, these MF equations of motion fail to accurately describe non-adiabatic processes in the regime of weak electronic coupling, due to the neglect of fluctuations in the electronic state variables.  The aim of the next section is thus to develop a continuous RPMD that preserves the kinetically important fluctuations in the electronic variables (\emph{i.e.}, ring-polymer ``kink-pair" formation).

\subsection{Kinetically constrained (KC) RPMD}
This section describes the central methodological contribution of the  paper.  We present a state-representation RPMD method that retains the robust features of the position-representation RPMD  while also including the kinetically important fluctuations in the electronic degrees of freedom.  
The development of this method involves three basic components, which are sequentially presented in the following subsections.  First, we introduce a continuous auxiliary variable, $y$, that reports on kink-formation in the ring polymer, and its associated effective potential.  Second, we introduce a kinetic constraint on the ring-polymer equilibrium distribution that inhibits the formation of instanton paths across non-degenerate double wells, thus correcting 
a known failure of instanton-based methods in the deep tunneling regime.
And third, we derive an appropriate mass for the auxiliary variable, $y$.

\subsubsection{A collective variable that reports on kinks}
\label{sec:kinkvar}

\begin{figure}
\includegraphics[scale=1.0]{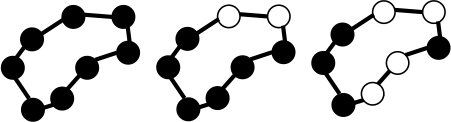} 
\caption{A schematic illustration of ring-polymer configurations that exhibit either zero (left), one (center), or two (right) kink-pairs.  Ring-polymer beads shown in white correspond to the electronic state $i^{(\alpha)}=0$, whereas those  in black correspond to $i^{(\alpha)}=1$.} 
\label{fig:rp} 
\end{figure}

The expression for the partition function in Eq. \ref{eq:standardPF} includes a sum
over the ensemble of ring-polymer configurations associated with all possible combinations of the electronic-state variables $\{i^{(\alpha)}\}$, namely
\begin{equation}
 \sum_{\{i^{(\alpha)}\}} \prod_{\alpha=1}^n M_{i^{(\alpha)},i^{(\alpha+1)}}(\rvec^{(\alpha)}).
\end{equation}
As is illustrated in Fig. \ref{fig:rp}, this ensemble includes  configurations 
for which all of the state variables assume the same value (i.e., 
$i^{(\alpha)}=0$ for all $\alpha$, or $i^{(\alpha)}=1$ for all $\alpha$), as well as ``kinked" ring-polymer configurations in which the electronic-state variable changes  value as a function of the bead index, $\alpha$. 
Because of the cyclic boundary condition for the ring-polymer coordinates, the number of kinks that is exhibited by a given configuration must be  even; we thus  refer to the number of ``kink-pairs" in describing the ring-polymer configuration.

The thermal weight of kinked ring-polymer configurations is closely related to the process of reactive tunneling. 
Indeed, for nuclear configurations in which the diabatic potentials are degenerate (i.e., $V_{0}(\rvec)=V_{1}(\rvec)$), 
the combined thermal weight of all ring-polymer configurations with $k$ kink-pairs is 
proportional to $(\beta K)^{2k}$. \cite{Cep95, Mar91,Kuk87}
This connection between imaginary-time path-integral statistics and the non-adiabatic coupling  $K$ lies at the heart of semiclassical instanton (SCI) theory,\cite{Cal83, Ben94,Cha75,Cal77,Han88,Mil75} 
and it underpins the accuracy of the RPMD method for the description of thermal reaction rates in the deep-tunneling regime.\cite{Ric09,Alt11,Alt11_2} 

For these reasons, 
the formation of kink-pairs during non-adiabatic transitions is an important feature to preserve in any extension of the RPMD method to multi-level systems. 
We thus introduce a discrete collective variable that reports on the existence of kink-pairs in the ring-polymer configuration, 
\begin{equation}
\theta(\{i^{(\alpha)}\})\!=\!\left\{
\begin{array}{c l}
-1,&i^{(\alpha)}=0\ \mbox{for all $\alpha$,}\\
1,&i^{(\alpha)}=1\ \mbox{for all $\alpha$,}\\
0, &\mbox{otherwise.}
\end{array}\right.
\label{collective}
\end{equation}
Furthermore, we introduce a continuous dummy variable $y$ that is tethered to $\theta(\{i^{(\alpha)}\})$ 
via a restraining potential $V_{\textrm{r}}(y, \{i^{(\alpha)}\})$, such that 
\begin{equation}
\label{eq:vrestraint}
e^{-\beta V_{\textrm{r}}(y, \{i^{(\alpha)}\})}  = f(y, \theta(\{i^{(\alpha)}\})),
\end{equation} 
where 
\begin{equation} 
f(y, \theta) = 
\lim_{b\rightarrow\infty} \frac{1}{2} 
\left(1-\tanh \left[b \left( |y-\theta | - \frac{1}{2} \right)\right]\right) .
\end{equation} 
Finally, the ring-polymer probability distribution in Eq. \ref{eq:standardPIdist} is reduced with respect to the discrete electronic variables $\{i^{(\alpha)}\}$,  yielding a distribution that depends only on the ring-polymer nuclear coordinates and on the  coordinate $y$ that smoothly reports on the existence of kink-pairs in the electronic coordinates, 
\begin{equation}
\label{eq:rho_rp}
 \rho_n(\allq,y)=
\Omega 
e^{-\beta V_{\mathrm{eff}}(\allq,y)},
\end{equation}
such that
\begin{equation}
\label{eq:rho_pf}
Z_n=\int\!\! d\allq \!\!\int\!\! dy\  \rho_n(\allq,y),
\end{equation}
and 
\begin{align}
\label{eq:veff}
V_{\mathrm{eff}}(&\allq,y)=U_{\mathrm{int}} (\allq) \\
&-\frac{1}{\beta}\ln \!  \left[ \sum _{\{i^{(\alpha)}\}}e^{-\beta V_{\textrm{r}}(y, \{i^{(\alpha)} \})} 
 \prod_{\alpha=1}^n M_{i^{(\alpha)},i^{(\alpha+1)}}(\rvec^{(\alpha)})\right].\nonumber  
\end{align}

Since $y$ is restrained to the collective variable $\theta(\{i^{(\alpha)}\})$, 
it is straightforward to obtain the free energy (FE)
of kink-pair formation via integration of  
 $\rho_n(\allq,y)$ over all values of $\allq$ and all values of $y$ that fall below a threshold magnitude, (i.e., $|y|<\epsilon$).   
 In practice, for a given number of ring-polymer beads $n$, the parameter $b$ is selected to be sufficiently large that this FE of kink-pair formation is invariant with respect to further increasing $b$.  This criterion leads to a well-defined limit for the convergence of both $n$ and $b$.

Note that the effective potential in Eq. \ref{eq:veff} introduces no approximation to the equilibrium statistics of the system; since the 
LHS of Eq. \ref{eq:vrestraint} is normalized  with respect to integration over $y$, then the expression for $Z_n$ in Eq. \ref{eq:rho_pf} 
is unchanged from Eq. \ref{eq:standardPF}.  
Eqs. \ref{eq:rho_rp} - \ref{eq:veff} thus correspond to a coarse-graining of the electronic degrees of freedom in a manner that is familiar from the 
description of large, purely classical systems \cite{Liw97,Izv05,Noi08,tfm07}
and that is not unlike the formulation of the centroid effective potential that appears in the centroid molecular dynamics (CMD) method for describing the quantized dynamics of nuclei.\cite{Cao94,Jan99}  
The auxiliary variable $y$ preserves key aspects of the fluctuations of the electronic coordinates by distinguishing kinked and unkinked ring-polymer configurations. 
As before, we can introduce classical equations of motion that rigorously preserve the quantized equilibrium distribution $\rho_n(\allq,y)$, namely 
\begin{equation}
\label{eq:rpmd_eoms1}
\begin{split}
\dot{v}_j^{(\alpha)}= &-\frac{n}{m_j} \frac{\partial}{\partial R_j^{(\alpha)}} V_{\mathrm{eff }}(\allq,y) \\
\dot{v}_{y}= &-\frac{1}{m_y}\frac{\partial}{\partial y} V_{\mathrm{eff }}(\allq,y),
\end{split}
\end{equation}
where we again utilize masses for the nuclear degrees of freedom that correspond to the physical masses of the system. 
We will shortly (in Subsection \ref{sec:mass}) introduce a criterion for the mass associated with auxiliary electronic variable, $m_y$.

The equations of motion in Eq. \ref{eq:rpmd_eoms1}, with an appropriate selection of $m_y$, fully specify an RPMD method for non-adiabatic systems that 
 preserves the exact  quantum Boltzmann distribution and that
explicitly accounts for  fluctuations in the electronic degrees of freedom.  
However,
 like the conventional position-representation RPMD method, these dynamics would overestimate ET rates in the Marcus inverted regime;\cite{Men11}  
to address this problem, the following subsection introduces a small modification to the quantized equilibrium distribution $\rho_n(\allq,y)$ that penalizes ring-polymer kink-pair formation between non-degenerate electronic states, 
thus yielding RPMD equations of motion that correctly describe non-adiabatic reactions across multiple regimes.

\subsubsection{A kinetic constraint on the quantum Boltzmann distribution}
\label{sec:kc}
Recent work 
 has established that many of the 
 successes and failures of the RPMD method in the deep tunneling regime arise from its close connection to semiclassical instanton theory.\cite{Ric09,Alt11,Alt11_2, Men11}  
 In a particularly striking failure of instanton-based methods, the rate of deep-tunneling across strongly asymmetric barriers is significantly overestimated in RPMD and steepest-descent SCI calculations, which manifests in incorrect rate coefficients for ET in the Marcus inverted regime.\cite{Men11, Shu13}  
A simple and methodologically suggestive way to understand this overestimation 
is to recognize that ring-polymer configurations associated with transitions between asymmetric potential wells (i.e., kinked ring-polymer configurations across non-degenerate diabatic surfaces, such that 
$|V_{0}(\rvec)-V_{1}(\rvec)| \gg |K(\rvec)|\ $)
appear with greater probability  in the equilibrium 
distribution than is appropriate for an accurate transition-state theory (TST) description of the deep-tunneling process.\cite{Men11}

To address this failure of instanton-based rate theories, 
we propose a simple modification of the 
path-integral distribution in Eq. \ref{eq:rho_rp} 
that explicitly penalizes 
the formation of kink-pairs at ring-polymer configurations for which the diabatic surfaces are non-degenerate, such that 
\begin{equation}
\begin{split}
\label{eq:rho_kc}
\rho_n^\mathrm{KC}(\allq,y)= \Omega e^{-\beta V_{\textrm{eff}}^{\mathrm{KC}}( \allq, y)},
\end{split}
\end{equation}
where
\begin{align*}
\label{eq:veff_kc}
&V_{\mathrm{eff}}^{\mathrm{KC}}(\allq,y)=U_{\mathrm{int}}(\allq)  \numberthis\\
& \quad\quad\quad -\frac{1}{\beta}\!\ln\!\!\left[ \!\sum _{\{i^{(\alpha)}\}} \!\!\!g(\{i^{(\alpha)}\}, \allq) \times \right. \nonumber\\ 
& \quad\quad\quad\quad\quad\quad  \left. e^{-\beta V_{\textrm{r}}(y, \{i^{(\alpha)} \})}   \!\!\prod_{\alpha=1}^n \! M_{i^{(\alpha)},i^{(\alpha+1)}}\!(\rvec^{(\alpha)})\!\right]\!\!,\nonumber  
  \end{align*}
and
 \begin{equation}
 \label{eq:g_func}
g(\{i^{(\alpha)}\},\allq)\!=\!\left\{
\begin{array}{c l}
1,&i^{(\alpha)}=0\ \mbox{for all $\alpha$,}\\
1,&i^{(\alpha)}=1\ \mbox{for all $\alpha$,}\\
\left( \frac{a}{\pi}\right)^{\frac{1}{2}}\eta e^{ - aw^2(\bar\rvec) }, &\mbox{otherwise.}
\end{array}\right.
\end{equation} 
The function $w(\rvec)=\left(V_{0}(\rvec)-V_{1}(\rvec)\right)/K(\rvec) $ 
 is the scaled difference in the diabatic potential surfaces, $\bar{\rvec}=\frac{1}{n}\sum_{\alpha=1}^n \rvec^{(\alpha)}$ is the ring-polymer centroid coordinate,  $a$ is a unitless convergence parameter, and 
\begin{equation}
\label{eq:eta}
\eta  = \langle |\nabla w(\rvec)| \rangle_\textrm{c}. 
\end{equation}
The brackets denote an ensemble average constrained to the intersection of the diabatic surfaces, 
such that
\begin{equation}
\label{eq:const_avg}
\langle (...) \rangle_\textrm{c}  =  \frac{ \int d\rvec \delta(w(\rvec)) (...) \left|K(\rvec)\right|^2 e^{-\beta V_{0}(\rvec) } }{ \int d \mathbf{R}  \delta(w(\rvec)) \left|K(\rvec)\right|^2 e^{-\beta V_{0}(\rvec)  }}.
\end{equation}
%
The exponential term in $g(\{i^{(\alpha)}\},\allq)$ penalizes the formation of ring-polymer kink-pairs 
as a function of the difference of the diabatic surfaces, and the associated prefactor ensures that the FE of kink-pair formation at the crossing of the diabatic surfaces is the same in the modified and unmodified distributions. 
In Appendix \ref{app:c_prefs}, we present the  detailed derivation of the penalty function $g(\{i^{(\alpha)}\},\allq)$; 
in Appendix \ref{app:bellalg}, we demonstrate that
 this form of the penalty function enables the effective potential in Eq. \ref {eq:veff_kc} and its derivatives to be factorized and efficiently evaluated in $\mathcal{O}(n)$ operations, 
 which is essential for  practical applications. 

A consequence of including the penalty function $g(\{i^{(\alpha)}\},\allq)$ is that the resulting  partition function
\begin{equation}
\label{eq:pi_kc}
Z_n^{\mathrm{KC}}=\int\!\! d\allq \!\!\int\!\! dy  \rho_n^{\mathrm{KC}}(\allq,y)
\end{equation}
is no longer identical to the result in Eq. \ref{eq:standardPF};  
the penalty function thus introduces an approximation to the true quantum Boltzmann statistics of the system.
However, two points are worth noting about this.  
Firstly, the configurations that are explicitly excluded via the penalty function constitute only a subset of those for which the ring polymer  exhibits kinks in the  electronic variables. 
If 
these excluded configurations are statistically unfavorable relative to unkinked configurations, which is generally true for cases in which the diabatic basis is a good representation for the electronic structure of a physical system, then
we may expect that the penalty function introduces little bias to the equilibrium properties of the system; 
regardless, the impact of the penalty function is easily tested by  sampling the path-integral statistics  both with and without this modification to the ring-polymer distribution. 
Secondly, we note that the ring-polymer configurations that are excluded via the penalty function are precisely those that give rise to 
the breakdown of the instanton approximation for tunneling across asymmetric barriers.  In this sense, we are introducing a  targeted kinetic constraint  on the accessible ring-polymer configurations with the aim of eliminating a known pathology of the semiclassical instanton theory upon which RPMD rests in the deep-tunneling regime.

The parameter $a$ in Eq. \ref{eq:g_func} dictates the strength of the  kinetic constraint that is introduced via the penalty function.
Convergence with respect to this parameter requires that the statistical weight of kinked ring-polymer configurations that violate the kinetic constraint must become negligible in comparison to the statistical weight of kinked configurations that satisfy the kinetic constraint.  We thus choose $a$ to be sufficiently large to converge the FE of kink-pair formation in the kinetically constrained ring-polymer distribution, which is given by $\Delta F^{\mathrm{KC}} = F^{\mathrm{KC}}(0)-F^{\mathrm{KC}}(-1)$, where 
\begin{equation}
F^{\mathrm{KC}}(y)=-\frac{1}{\beta}\ \textrm{ln} 
\int\!\! d\allq \rho_n^{\mathrm{KC}}(\allq,y).
\end{equation}
This criterion provides a simple basis for the determination of $a$ in a given application.
 However, it should also be noted that if $a$ is chosen to be 
 greater than unity, then  kink-pair formation will be hindered at ring-polymer configurations for which $|V_{0}(\rvec)-V_{1}(\rvec)| < |K(\rvec)|$.  
 Therefore, in addition to requiring that $a$ be sufficiently large to converge the FE of kink-pair formation in the kinetically constrained ring-polymer distribution, we also require that the parameter 
not exceed a value of unity.  In principle, systems for which this range of convergence does not exist fall outside the realm of applicability of the current method and are likely to be better 
described using the MF non-adiabatic RPMD in Eq. \ref{eq:mf_eoms}.  However, all of the 
systems considered in the current paper exhibit this range of convergence with 
$a <1$, suggesting that the existence of a range of convergence for this parameter 
is a relatively minor concern.

The classical equations of motion associated with the equilibrium distribution $\rho_n^{\mathrm{KC}}(\allq,y)$  are 
\begin{equation}
\begin{split}
\label{eq:rpmd_eoms}
\dot{v}_j^{(\alpha)}= &-\frac{n}{m_j} \frac{\partial}{\partial R_j^{(\alpha)}} V_{\mathrm{eff }}^{\mathrm{KC}}(\allq,y)  \\
\dot{v}_{y}= &-\frac{1}{m_y}\frac{\partial}{\partial y} V_{\mathrm{eff }}^{\mathrm{KC}}(\allq,y). 
\end{split}
\end{equation}
Eq. \ref{eq:rpmd_eoms}  specifies the kinetically constrained RPMD (KC-RPMD)   method for non-adiabatic dynamics, which explicitly accounts for  fluctuations in the electronic degrees of freedom and which addresses the 
failing of instanton-based methods in describing  deep-tunneling across asymmetric barriers.
As before, these equations utilize the physical masses for the nuclear degrees of freedom, and  $m_y$ will be described in the following Subsection \ref{sec:mass}.

We emphasize that since the trajectories generated by  Eq. \ref{eq:rpmd_eoms} rigorously 
preserve a well-defined (albeit approximate) equilibrium distribution, 
the KC-RPMD method exhibits all of the robust features of the usual position-representation RPMD method, including detailed balance, 
time-reversibility, invariance of thermal rate coefficient calculations to the choice of dividing surface, and the ability to immediately utilize the full machinery of classical MD simulations.\cite{tfm2013} 
However, unlike the  position-representation RPMD method,  KC-RPMD allows for the description of non-adiabatic processes involving many-electron wavefunctions and will be shown to overcome the previous failures of instanton-based methods for ET reactions in the Marcus inverted regime.

\subsubsection{The mass of the auxiliary variable} 
\label{sec:mass}

For the  position-representation RPMD method,\cite{Cra04,tfm2013}  
the correspondence between the ring-polymer bead masses and the physical masses of the particles in the system has been justified in several ways.  These  include the demonstration that the RPMD mass choice 
leads to both \emph{(i)} optimal agreement in the short-time limit between general, real-time quantum mechanical correlation functions and their RPMD approximations\cite{Bra06} and
\emph{(ii)} an RPMD TST that corresponds to the $t\to0^+$ limit of an appropriately transformed quantum-mechanical flux-side correlation function, and therefore yields the exact quantum rate coefficient in the absence of recrossing.\cite{Ric09, Hel13, Hel13_2}

In the current study, we employ a justification similar to \emph{(ii)} for the determination of 
$m_y$, the mass of the auxiliary variable that reports on ring-polymer kink formation. 
Specifically, we choose $m_y$ such that 
the resulting KC-RPMD TST exactly recovers the multi-dimensional Landau-Zener TST rate expression for non-adiabatic transitions in the weak-coupling regime.\cite{Sti76} 
The resulting expression, which is derived in Appendix \ref{app:mass}, is
\begin{equation}
\label{eq:mass_my}
m_y=\frac{\beta^3 \hbar^2}{2 \pi^3}\left[  \frac{ \langle |\nabla w(  \rvec)| \rangle_\textrm{c}}{\langle | K( \rvec )|^{-1} \rangle_\textrm{c}} \right ]^2,
\end{equation}
where the constrained ensemble average is defined in Eq. \ref {eq:const_avg}.
For simple potentials, this expression can be evaluated analytically; however,  for general systems, the evaluation of $m_y$ involves only a constrained  ensemble average, which 
 can be performed using well-established classical simulation methods\cite{Frenkel}
and which is already required for most RPMD 
(or classical mechanical) rate calculations.\cite{tfm2013} 

\subsubsection{Summary of the KC-RPMD method}

Before proceeding, we  summarize the steps that are needed to implement the KC-RPMD method for a given application, which emphasizes the relative simplicity  
of this non-adiabatic extension of RPMD.  

\begin{enumerate}
\item Determine the number of ring-polymer beads, $n$, needed to converge the equilibrium properties of the system in the path-integral representation, as is typically necessary in  path-integral calculations.
\item Converge the coefficient $b$ that appears in the potential of restraint (Eq. \ref{eq:vrestraint}) between the auxiliary variable $y$ and the collective variable that reports on the existence of kinks in the ring-polymer configuration.  As is described in Subsection \ref{sec:kinkvar}, the coefficient $b$ should be sufficiently large to converge the FE of kink-pair formation $\Delta F^{\mathrm{KC}}$. 
\item Compute the mass $m_y$ (Eq. \ref{eq:mass_my}) and $\eta$ (Eq. \ref{eq:eta}) from a single, constrained ensemble average.
\item Converge the coefficient $a$ that appears in the function  that penalizes the weight of 
kinked ring-polymer configurations across non-degenerate diabatic surfaces  (Eq. \ref{eq:g_func}). As is described in Subsection \ref{sec:kc}, the coefficient $a$ should be sufficiently large to converge $\Delta F^{\mathrm{KC}}$ but should not exceed a value of unity.
\item As for the usual position-representation RPMD method, model the real-time dynamics of the system by integrating classical equations of motion in an extended phase space, as defined by Eq. \ref{eq:rpmd_eoms}.
\end{enumerate}

\section{Model Systems}
\label{sec:systems}
Numerical results 
are presented for model systems with potential energy functions of the form
 \begin{equation}
\hat{V}(\rvec) = \hat{V}_\mathrm{S}(\rvec) +\mathds{1} V_\mathrm{B}(\rvec),
\end{equation}
where $\mathds{1}$ is the identity operator,
\begin{equation}
\hat{V_\mathrm{S}}(\rvec) = 
\begin{pmatrix}
V_{0}(s) &  K \\
K & V_{1}(s)  \end{pmatrix},
\end{equation}
$K$ is a constant, 
$s$ is a one-dimensional (1D) system coordinate,
and the full set of nuclear position coordinates $\rvec=\{s,\xvec\}$ includes 
a set of 
$f$  bath modes, $\xvec$. We use atomic units throughout, unless otherwise noted. 

System A models a simple avoided-crossing reaction in the absence of a dissipative bath, for which 
\begin{equation}
\label{eq:sysa_pot}
\hat{V_\mathrm{S}}(s) = 
\begin{pmatrix}
A e^{Bs}&  K \\
K& A e^{-Bs}  \end{pmatrix}
\end{equation}
and $V_\mathrm{B}(\rvec)=0$.  
Parameters for this model are presented in Table \ref{tab:sysa}, and the quantities $\eta$ and $m_y$ are  analytically evaluated from Eqs. \ref{eq:eta} and \ref{eq:mass_my}, such that 
$\eta=8\times10^{-2}$ 
and the values for $m_y$ 
are given in Table \ref{tab:sysa_my}.

System B models a condensed-phase ET reaction in various regimes, with the redox system described using 
\begin{equation}
 \hat{V}_\mathrm{S}(\rvec) = 
\begin{pmatrix}
As^2+Bs&  K \\
K& As^2-Bs+\epsilon  \end{pmatrix},
\end{equation}
where $s$ corresponds to the  local solvent dipole. 
This solvent coordinate is  linearly coupled to a bath of 
harmonic  oscillators, such that 
\begin{eqnarray}
\label{eq:u_sysbath}
V_{\mathrm{B}}(s,\mathbf{x})&=&\sum_{j=1}^f  \left[ \frac{1}{2} M\omega_j^2 \left(x_j-\frac{c_j s}{M \omega_j^2} \right)^2\right],
\end{eqnarray}
with oscillators of mass $M$.
The bath exhibits an Ohmic spectral density with cutoff frequency $\omega_{\mathrm{c}}$,
\begin{equation}
\label{eq:spectraldensity}
J(\omega)=\gamma \omega e^{-\omega/\omega_{\mathrm{c}}},
\end{equation}
where $\gamma$ is a dimensionless parameter that controls the strength of coupling between the system and the bath modes and that is chosen to be characteristic of a condensed-phase environment. The spectral density in Eq. \ref{eq:spectraldensity} is discretized into $f$ oscillators with frequencies \cite{Cra05}
\begin{equation}
\omega_j=-\omega_{\mathrm{c}}\ln\left(\frac{j-0.5}{f}\right)
\end{equation}
and coupling constants 
\begin{equation}
c_j=\omega_j\left(\frac{2\gamma M \omega_{\mathrm{c}}}{f\pi}\right)^{1/2},
\end{equation}
where $j=1\ldots f$.
The additional parameters for System B are provided in Table \ref{tab:sysb}, and  $m_y=3.94\times10^{4}$ and $\eta=6.86\times10^4$ are again  evaluated from Eqs. \ref{eq:eta} and \ref{eq:mass_my}.

In the following, we consider examples in which the system coordinate $s$ is either quantized or treated in the classical limit.
However, to enable  straightforward comparison with other methods, we will in all cases consider the classical limit for the nuclear degrees of freedom associated with the harmonic oscillator bath.  
As is usual for applications of RPMD,\cite{tfm2013}
the classical limit for nuclear degrees of freedom is obtained by requiring the associated
ring-polymer bead positions 
to coincide.

\begin{table}
\caption{Parameters for System A.\footnote{Unless otherwise noted, values are reported in atomic units.}
\label{tab:sysa}}
\begin{ruledtabular}
\begin{tabular}{ccccc}
 Parameter & Value Range \\
\hline
$A$ & $0.02$ \\ 
$B$ & $2.0$ \\
$K$ & $5\times 10^{-5}$  \\
$m_{s}$ & $2000$  \\
$1000/T($K$)$ & $1.5 -  5.5$    \\
\end{tabular}
\end{ruledtabular}
\end{table}

\begin{table}
\caption{Values of $m_y$ for the KC-PMD simulations of System A.\footnote{Unless otherwise noted, values are reported in atomic units.}
\label{tab:sysa_my}}
\begin{ruledtabular}
\begin{tabular}{ c c }
$1000/T($K$)$  & $m_\mathrm{y}$ \\
\hline
$1.5$ & $2.74 \times 10^3$ \\
$2.0$ & $6.50 \times 10^3$ \\
$2.5$ & $1.27 \times 10^4$ \\
$3.0$ & $2.19 \times 10^4$ \\
$3.5$ & $3.48 \times 10^4$ \\
$4.0$ & $5.20 \times 10^4$ \\
$4.5$ & $7.40 \times 10^4$ \\
$5.0$ & $1.02 \times 10^5$ \\
$5.5$ & $1.35 \times 10^5$ \\
\end{tabular}
\end{ruledtabular}
\end{table}

\begin{table}
\caption{Parameters for System B.\footnote{Unless otherwise noted, values are reported in atomic units.}
\label{tab:sysb}}
\begin{ruledtabular}
\begin{tabular}{ccccc}
 Parameter & Value Range \\
\hline
$A$ & $4.772 \times 10^{-3}$ \\ 
$B$ & $2.288 \times 10^{-2}$ \\
$\epsilon$ & $0 - 0.236$ \\
$K$ & $6.67\times10^{-7} - 7.5\times10^{-3}$  \\
$m_{s}$ & $1836.0$  \\
$M$  &   $1836.0$ \\
$\omega_{\mathrm{c}} $ & $2.288 \times 10^{-2}$  \\
$\gamma/ M \omega_{\mathrm{c}} $ &  $1.0$ \\
 $f$ &  $12$     \\
$T$ & $300$ K  \\
\end{tabular}
\end{ruledtabular}
\end{table}

\section{Calculation of  reaction rates}
\label{sec:kcrpmd_rates}

\subsection{Calculation of  KC-RPMD rates}
As for the position-representation RPMD method,\cite{tfm2013} the KC-RPMD method involves classical equations of motion in an extended phase space (Eq. \ref{eq:rpmd_eoms}).
%
%
 Accordingly, standard methods for the calculation of classical reaction rates can be employed to compute KC-RPMD reaction rate coefficients,\cite{Frenkel} and 
 the KC-RPMD rate can be separated into statistical and dynamical contributions as\cite{Cha78, Ben77}
 %
\begin{equation}
\label{eq:kcrpmd_rate}
k^\mathrm{KC-RPMD}=k_{\mathrm{TST}}^{\mathrm{KC-RPMD}} \lim_{t\to\infty}\kappa(t),
\end{equation}
where $k_{\mathrm{TST}}^{\mathrm{KC-RPMD}}$ is the TST estimate for the rate associated with the dividing surface $\xi(\mathbf{r})=\xi^\ddagger$, and $\kappa(t)$ is the time-dependent transmission coefficient that corrects for  dynamical recrossing at the dividing surface.  Here, $\xi(\mathbf{r})$ is a collective variable that distinguishes between reactant and product basins of stability, defined as a function of the position vector of the full system in the ring-polymer representation, $\mathbf{r}=\left\{{\allq,y}\right\}$.  

The KC-RPMD TST rate is calculated using the usual 
 expression,\cite{tfm2013}  
\begin{equation}
\label{eq:kcrpmd_tst_def}
k_{\mathrm{TST}}^{\mathrm{KC-RPMD}}=\frac{1}{\sqrt{2 \pi \beta }} \langle{\chi_\xi}\rangle^{\ddagger} \frac{e^{-\beta \Delta F(\xi^\ddagger)}}{\int_{-\infty}^{\xi^\ddagger} d\xi e^{-\beta \Delta F(\xi)} }.
\end{equation}
Here, $F(\xi)$ is the FE along $\xi$ relative to a reference value $\xi^\circ$, such that
\begin{equation}
e^{-\beta \Delta F(\xi^\ddagger)}=\frac{\langle\delta (\xi(\mathbf{r})-\xi^\ddagger)\rangle}{\langle\delta (\xi(\mathbf{r})-\xi^\circ)\rangle},
\end{equation}
and \cite{Car89,Sch03,Wat06} 
\begin{equation}
\label{eq:geom_factor}
\chi_\xi(\mathbf{r}) = \left[ \sum_{j}^{nd+1} \frac{1}{m_j} \left(\frac{\partial \xi (\mathbf{r})}{\partial r_j} \right)^2 \right]^{1/2}.
\end{equation}
The sum in Eq. \ref{eq:geom_factor} runs over all the $nd+1$ 
degrees of freedom for the ring-polymer representation used here, and $m_j$ denotes the mass associated with each degree of freedom. The angle brackets indicate an equilibrium ensemble average 
\begin{equation}
\langle \dots \rangle = \frac {\int d \mathbf r \int d\mathbf{v}  \; e^{-\beta H(\mathbf{r},\mathbf{v})} (\dots)}{\int d \mathbf r \int d\mathbf{v} \; e^{-\beta H(\mathbf{r},\mathbf{v})} },
\end{equation} 
where $\mathbf{v}=\left\{{ \left\{ 
v^{(\alpha)} \right\} ,v_y}\right\}$ is the velocity vector for the full system in the ring-polymer representation and $H(\mathbf{r},\mathbf{v})$ is the ring-polymer Hamiltonian associated with the KC-RPMD effective potential. 
 Similarly, 
 \begin{equation}
\langle \dots \rangle^\ddagger = \frac {\int d \mathbf r \int d\mathbf{v} \; e^{-\beta H(\mathbf{r},\mathbf{v})} \delta (\xi(\mathbf{r})-\xi^\ddagger)(\dots)}{\int d \mathbf r \int d\mathbf{v} \; e^{-\beta H(\mathbf{r},\mathbf{v})} \delta (\xi(\mathbf{r})-\xi^\ddagger)}
\end{equation} 
is the ensemble average 
constrained to the dividing surface.
For the case of $\xi(\mathbf{r})=y$, the KC-RPMD TST rate expression takes a particularly concise form,
\begin{equation}
\label{eq:kcrpmd_tst_y}
k_{\mathrm{TST}}^{\mathrm{KC-RPMD}}=\frac{1}{\sqrt{2 \pi \beta m_y }} \frac{e^{-\beta \Delta F(y^\ddagger)}}{\int_{-\infty}^{y^\ddagger} dy e^{-\beta \Delta F(y)} }.
\end{equation}  
The  transmission coefficient in Eq. \ref{eq:kcrpmd_rate} is calculated as
\begin{equation}
\label{eq:kcrpmd_kappa}
\kappa(t)=\frac{\langle \dot\xi_0 h\left(\xi(\mathbf{r}_t)-\xi^\ddagger\right) \rangle^\ddagger}{\langle \dot \xi_0 h (\dot \xi_0) \rangle^\ddagger},
\end{equation}
where $h(x)$ is the Heaviside  function, and the subscripts 0 and $t$ denote evaluation of the quantity from the trajectory at its initiation and after evolution for time $t$, respectively. 


\subsubsection{KC-RPMD rate calculation in System B}

The KC-RPMD reaction rate for System B is calculated as the product of the KC-RPMD TST rate (Eq. \ref{eq:kcrpmd_tst_y}) and the transmission coefficient (Eq. \ref{eq:kcrpmd_kappa}).  In all cases, the TST dividing surface is defined as an isosurface of the auxiliary variable, $y$.

We perform two sets of KC-RPMD reaction rate calculations for System B. In the first, the non-adiabatic coupling $K = 6.67\times10^{-7}$ is held fixed, $T=300$ K, and the driving force parameter $\epsilon$ is varied. The ring polymer is discretized using $n=32$ beads.  For cases in which the solvent dipole coordinate $s$ is treated classically, the ring-polymer  bead positions for this solvent  coordinate 
are restricted to coincide; in all cases, the degrees of freedom associated with the harmonic oscillator bath are treated classically.
Convergence checks with respect to the strength of the kinetic constraint, $a$, are provided in the Results Section.  Unless otherwise stated, the results for this set of calculations are reported using $a=5\times 10^{-8}$.

The KC-RPMD TST rate (Eq. \ref{eq:kcrpmd_tst_y}) is obtained from $F(y)$, the FE profile in the continuous auxiliary variable.  For cases in which the solvent coordinate $s$ is treated classically, 
the FE profile is obtained by direct numerical integration; 
for cases in which the solvent coordinate is quantized, 
the FE profile is calculated using umbrella sampling and the weighted histogram analysis method (WHAM).\cite{Frenkel,Kum92,Kum94,Rou95} In the latter case, for each value of $\epsilon$,  $F(y)$ 
is obtained by reducing the two-dimensional (2D) FE surface computed with respect to $y$ and the ring-polymer centroid for the solvent coordinate,  $\bar{s}$.

The 2D FE profile $F(\bar{s},y)$ is sampled using independent KC-RPMD trajectories with a  potential that restrains $\bar {s}$ and $y$ to  $s_0$ and $y_0$, respectively, such that 
\begin{align*}
V_{\mathrm{map}}&\left(\{s^{(\alpha)}\},y\right)=V_\mathrm{eff}^{\mathrm{KC}}\left(\{s^{(\alpha)}\},y\right)+\numberthis\\
&0.5k_s(\bar{s}-s_0)^2 + \left ( 0.5k_y(y-y_0)^2+ 10k_y(y-y_0)^6 \right).
\end{align*}
The KC-RPMD sampling trajectories are grouped into two sets.
 The first set is comprised of 1100  trajectories 
 that primarily sample the reactant and product basins, with  $s_0$ and $y_0$ assuming values on a square grid. 
The parameter $s_0$ assumes $22$ uniformly spaced values in the region $s_0=[-4,9]$, and the associated force constant is $k_s=0.04$. 
For each value of $s_0$, the parameter $y_0$ assumes 
$10$ equally-spaced values in the range $y_0=[-1.5,-0.5]$ with  $k_y=0.2$,
$10$ equally-spaced values in the range $y_0=[1.5,0.5]$ with  $k_y=0.2$, 
$15$ equally-spaced values in the range $y_0=[-0.5,-0.2]$ with $k_y=16.0$,
and
$15$ equally-spaced values in the range $y_0=[0.5,0.2]$  with $k_y=16.0$. 
 The second set of sampling trajectories is comprised of 506 KC-RPMD trajectories 
 that primarily sample the region of the intersection of the diabatic surfaces, denoted $s^\ddagger$, with  $s_0$ and $y_0$ assuming values on a square grid. 
 The parameter $s_0$ assumes $11$ uniformly spaced values in the region $s_0=[s^\ddagger-0.2, s^\ddagger+0.2]$, and the associated force constant is $k_s=4.0$. 
For each value of $s_0$, the parameter $y_0$ assumes 
$13$ equally-spaced values in the range $y_0=[0.40, 0.52]$ with  $k_y=64.0$,
$13$ equally-spaced values in the range $y_0=[-0.40, -0.52]$ with  $k_y=64.0$, 
and 
$20$ equally-spaced values in the range $y_0=[-0.4, 0.4]$ with $k_y=6.0$.
 Each sampling trajectory is evolved for at least $20$ ps using a timestep of  $dt=0.02$ fs.  Thermostatting is performed by re-sampling the velocities from the Maxwell-Boltzmann (MB) distribution every $200$ fs. 

The transmission coefficients (Eq. \ref{eq:kcrpmd_kappa}) are calculated using KC-RPMD trajectories that are released from the dividing surface associated with $y^\ddagger=0$.  For each value of the driving force $\epsilon$, a total of $1000$ trajectories are released.  
Each KC-RPMD trajectory is evolved for $200$ fs using a timestep of $dt=0.02$ fs and with the initial velocities sampled from the MB distribution.
The initial configurations for the KC-RPMD trajectories are generated from long KC-RPMD trajectories that are constrained to the dividing surface using the RATTLE algorithm;\cite{And83}
the constrained trajectories are at least $200$ ps  in time and are thermostatted by resampling the velocities from the MB distribution every $200$ fs.

In the second set of KC-RPMD reaction rate calculations for System B, 
 $\epsilon=0$, $T=300$ K, and the non-adiabatic coupling $K$ is varied from the weak-coupling to the strong-coupling regimes, such that  $- \log(K)\in\{6.18, 6.00, 5.50, 5.00, 4.50, 4.00, 3.30, 3.00, 2.70, 2.30, 2.10\}$. 
For these couplings, the  calculations are performed using $-\log(a)\in\{7.3, 5.0, 4.0, 3.0, 2.0, 2.0, 1.5, 1.0, 0.5, 0.5, 0.5\}$, 
 respectively.  At each coupling, it is confirmed that the FE barrier in $F(y)$ and   the KC-RPMD rate are robust with respect to increasing the convergence parameter $a$, although at larger couplings, the plateau range for $a$ becomes more narrow. 
 The ring-polymer is discretized using $n=128$  beads, which is sufficient for convergence at all values of the non-adiabatic coupling; the solvent coordinate and the harmonic bath are treated classically.


\subsubsection{KC-RPMD rate calculation in System A} 
The form of the potential energy surface in System A precludes the use of the factorization shown in Eq. \ref{eq:kcrpmd_rate}, which assumes that the reactant and product basins are bound. The KC-RPMD rate in System A is instead evaluated directly as the long-time limit of the flux-side correlation function,
\begin{equation}
k^{\mathrm{KC-RPMD}}=\frac{1}{Q_\mathrm{R} (T)}\lim_{t \to \infty} C_{\mathrm{fs}}(t),
\end{equation}
where 
\begin{equation}
C_{\mathrm{fs}}(t) \!=\!\Omega \!\! \int  \!\! d \mathbf{r}_0  \!\! \int  \! \!d  \mathbf{v}_0 e^{-\beta H(\rv, \vv)} \delta (y_0)  v_y h (y_t).
\end{equation}
Here,  $\rv=\left\{\{s^{(\alpha)}\},y\right\}$, $\mathbf{v}=\left\{\{v^{(\alpha)}\},v_y\right\}$, and the subscripts denote the values of the ring-polymer positions and velocities at times $0$ and $t$, respectively. The reactant partition function for the unbound system is the inverse de Broglie thermal wavelength, $Q_\mathrm{R} (T)=
\sqrt{\frac{m_s}{2 \pi \beta \hbar^2}}$, and 
\begin{equation}
\Omega = \left(\frac{m_\mathrm{s}}{2 \pi \hbar }\right)^n \sqrt{\frac{m_\mathrm{y}\beta}{2 \pi}}.
\end{equation} 
Efficient Monte Carlo sampling of the initial conditions in the flux-side correlation function is accomplished by introducing two reference distributions, 
\begin{equation}
\rho^{\mathrm{ref}}_+(\rv,\vv) = e^{-\beta H_{\mathrm{ref}} (\rv,\vv)}  \delta (y)  h (v_y) v_y  
\label{eq:refdist1}
\end{equation}
and
\begin{equation}
\rho^{\mathrm{ref}}_-(\rv,\vv)=e^{-\beta H_{\mathrm{ref}} (\rv,\vv)}  \delta (y)  h (-v_y) v_y,
\label{eq:refdist2}
\end{equation}
where
\begin{align*}
 H_{\mathrm{ref}} (\rv,\vv)  \!=\!\sum_{\alpha=1}^n & \frac{1}{2} \tilde{m}_\mathrm{s} {v^{(\alpha)}}^2+\frac{1}{2} m_y v_y^2+ \numberthis \\
&U_{\mathrm{int}}(\{s^{(\alpha)}\})+V_{\mathrm{ref}}(\bar{s})
\end{align*} and 
\begin{equation}
V_{\mathrm{ref}}(\bar{s}) = - \frac {\bar{s}^2}{\sigma^2}.
\end{equation}
The difference between the reference and system Hamiltonians is thus given by
\begin{equation}
\Delta V(\rv,\vv)=H(\rv,\vv)  -H_{\mathrm{ref}}(\rv,\vv).
\end{equation}
The KC-RPMD rate is then evaluated using
\begin{align*}
&k^{\mathrm{KC-RPMD}}(T)  =  \lim_{ t \to \infty } \frac{\Omega}{Q_\mathrm{R} (T)} \times \numberthis\\
 &\!\! \left[ \Phi_+\!\! \left\langle e^{-\beta \Delta V(\rv_0,\vv_0)}  h (y_t) \right\rangle_+ + \Phi_- \!\! \left \langle e^{-\beta \Delta V(\rv_0,\vv_0)}  h (y_t) \right \rangle_- \right ],
\end{align*}
where the angle brackets denote sampling over the initial positions and velocities of the system using the distributions described by Eqs. \ref{eq:refdist1} and \ref{eq:refdist2},
\begin{equation}
\label{eq:sampdist}
\langle(\dots)\rangle_\pm = \frac{ \int  \! d\rv_0 \int d\vv_0 \ (\dots) \ \rho^{\mathrm{ref}}_\pm( \rv_0 ,\vv_0) }{ \int  d\rv_0 \int d\vv_0 \ \rho^{\mathrm{ref}}_\pm  ( \rv_0 ,\vv_0)},
\end{equation}
and $\Phi_\pm$ denote the value of the reference distributions integrated over all  space, 
\begin{equation}
\label{eq:phi_pm}
\Phi_\pm = \!\! \int  \!\! d \rv_0 \int \!d\pv_0 \;\rho^{\mathrm{ref}}_\pm ( \rv_0 ,\vv_0).
\end{equation}
The reference distributions involve integration over separable degrees of freedom, and Eq. \ref{eq:phi_pm} can be evaluated analytically.

For each temperature $T$, $2\times10^5$ initial configurations are sampled from the distribution in Eq. \ref{eq:sampdist}, and KC-RPMD trajectories are evolved for $500$ fs with a timestep of $d t=0.02$ fs. 
We employ $n=64$ ring-polymer beads and $a=5 \times 10^{-6}$; it is confirmed that 
varying $a$ over two orders of magnitude leads to graphically indistinguishable differences in the results.

\subsection{Calculation of reference TST rate expressions}

The exact quantum-mechanical thermal rate coefficient 
 for System A is 
\begin{equation}
\label{ExactQuantum}
k^{\mathrm{ex}}(T)=\frac{1}{Q_{\mathrm{R}}(T)}\frac{1}{2\pi\hbar} \int _0^{\infty} dE e^{-\beta E} N(E), 
\end{equation}
where $N(E)$ denotes the microcanonical reaction probability at energy $E$. These probabilities are evaluated directly by solving the scattering problem for the potential in Eq. \ref{eq:sysa_pot} using the log-derivative method.\cite{Man86,Joh73}

Reference values for the  thermal reaction rates for System B are evaluated using rate expressions for adiabatic and non-adiabatic ET.
The TST expression for adiabatic ET with classical solvent is \cite{Hush60, Mar85}  
\begin{equation}
\label{eq:kad}
k_{\mathrm{ET}}^{\mathrm{ad}}=\frac{\omega_\mathrm{s}}{2 \pi} \exp{[-\beta G^\ddagger_{\mathrm{ad}}]},
\end{equation}
where $\omega_\mathrm{s}$ and $G^\ddagger_{\mathrm{ad}}$ are respectively the solvent frequency and the FE barrier to reaction, calculated along the solvent coordinate. 
The expression for non-adiabatic ET with classical solvent is given by the classical Marcus Theory (MT) expression\cite{Mar85}
\begin{equation}
\label{eq:knad}
k_{\mathrm{ET}}^{\mathrm{nad}}=\frac{2\pi}{\hbar} |K|^2 \sqrt{\frac{\beta}{4\pi \lambda}}\exp{\left [-\beta \frac{(\lambda+\Delta G^o)^2}{4\lambda}\right]},
\end{equation}
where $\lambda$, $\Delta G^\circ$, and $K$ are the solvent reorganization energy, the driving force, and the electronic coupling, respectively.
The expression for non-adiabatic ET with quantized solvent is given by the golden-rule expression\cite{Uls75, Ulstrupbook, Mar85} 
\begin{equation}
\label{eq:qmt}
k_{\mathrm{ET}}^{\mathrm{nad}}\!=\!\frac{2\pi}{\hbar Q_{\mathrm{R}}} |K|^2 \sum_\mu\! \sum_\nu e^{-\beta E^{(\textrm{a})}_{\mu}} |\braket{\chi_\mu}{\chi_\nu}|^2 \delta(E^{(\textrm{a})}_{\mu}-E^{(\textrm{b})}_{\nu}),
\end{equation}
where $\chi_{\mu}$ and $\chi_\nu$ denote the reactant and product vibrational eigenstates for the solvent coordinate, respectively, with associated energies $E^{(\textrm{a})}_{\mu}$ and $E^{(\textrm{b})}_{\nu}$. If the reactant and product solvent potential energy  surfaces are represented by displaced harmonic oscillators with frequency $\omega_\mathrm{s}$, as is the case for System B,  this equation can be transformed into the analytical form,\cite{Uls75, Ulstrupbook} 
\begin{equation}
\label{eq:qmt2}
k_{\mathrm{ET}}^{\mathrm{nad}}=\frac{2\pi}{\hbar \omega_\mathrm{s}} |K|^2 e^{v z - S \coth(z)} I_{v}(S\; \mathrm{csch}(z)),
\end{equation}
where $z= { \beta \omega_\mathrm{s}}/{2}$, $v=-{\epsilon}/{\omega_\mathrm{s}}$, $I_{v}$ is a modified Bessel function of the first kind, and $S=(2\hbar)^{-1}{m_\mathrm{s} \omega_\mathrm{s}}  {\Delta_\mathrm{s}}^2 $, with $\Delta_\mathrm{s}$  and $\epsilon$ denoting the relative horizontal displacement of the diabatic potential energy surfaces and the reaction driving force, 
respectively.

\section{Results}
\label{sec:results}
We  present numerical results obtained using the new KC-RPMD method, including 
comparisons with reaction rates obtained using exact quantum mechanics 
(Eq. \ref{ExactQuantum}), position-representation RPMD, MF non-adiabatic RPMD (Eq. \ref{eq:mf_eoms}), and TST rate expressions (Eqs. \ref{eq:kad}-\ref{eq:qmt}). These results demonstrate the performance of the KC-RPMD method in models for 
a simple avoided-crossing reaction and for condensed-phase ET. 
We examine these models in a variety of regimes to demonstrate the performance of the KC-RPMD in describing electronically adiabatic vs. non-adiabatic reactions, classical vs. quantized nuclei, and  normal vs. inverted ET.

\subsection{Simple avoided-crossing reaction}

\begin{figure}[h]
\includegraphics{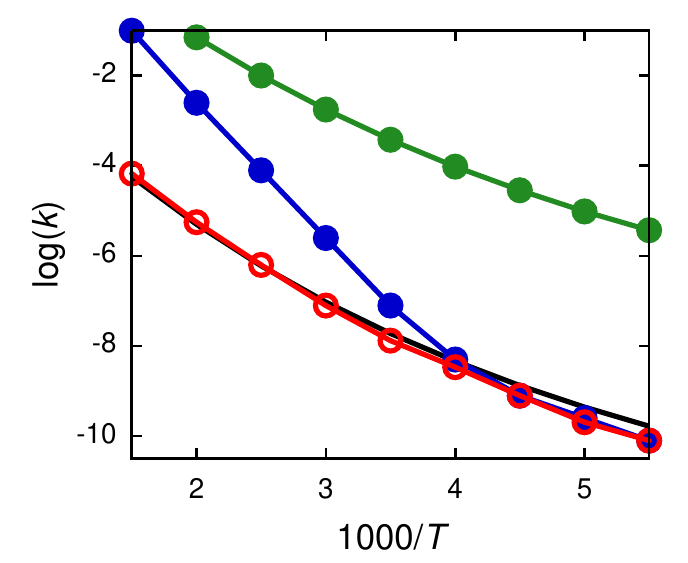}
\caption{ 
Thermal reaction rate coefficients  for System A 
%
as a function of temperature, obtained using KC-RPMD (red), MF non-adiabatic RPMD (blue), position-representation RPMD on the lower adiabatic surface (green), and exact quantum mechanics (black).}
\label{fig:dexp} 
\end{figure}

We begin by considering numerical results for System A, which models a non-dissipative avoided-crossing reaction in 1D.
Figure \ref{fig:dexp} presents the thermal reaction rate for this system over the range of temperatures from $187$ to $667$ K,  which corresponds to spanning from the weak- to moderate-coupling regimes (i.e., $\beta K=0.02-0.1$). 
The reaction rates are computing using the KC-RPMD and MF non-adiabatic RPMD methods. For comparison, we also include the rates calculated with position-representation RPMD on the lower adiabatic surface, and exact rates computed using the log-derivative method. 

Comparison of the position-representation RPMD rates and the exact quantum rates illustrate the importance of non-adiabatic effects in this model.   The MF non-adiabatic RPMD method, which incorporates non-adiabatic effects via the thermal average of fluctuations in the electronic degrees of freedom, does well in regimes of stronger coupling but breaks down when the statistical weight of ring-polymer configurations with kink-pairs becomes small relative to the weight of configurations without kink-pairs. In contrast, KC-RPMD performs well throughout the entire range of temperatures, accurately capturing the regime for which the mean-field result is accurate as well as the weak-coupling regime for which explicit fluctuations in the electronic degrees of freedom are important.

\subsection{Condensed-phase electron transfer}

We next present numerical results for System B, a system-bath model for condensed-phase ET. We consider the effects of varying the non-adiabatic coupling, changing the driving force, and including quantum-mechanical effects in the treatment of the solvent coordinate.  

\begin{figure}[h!]
\includegraphics[scale=1.0]{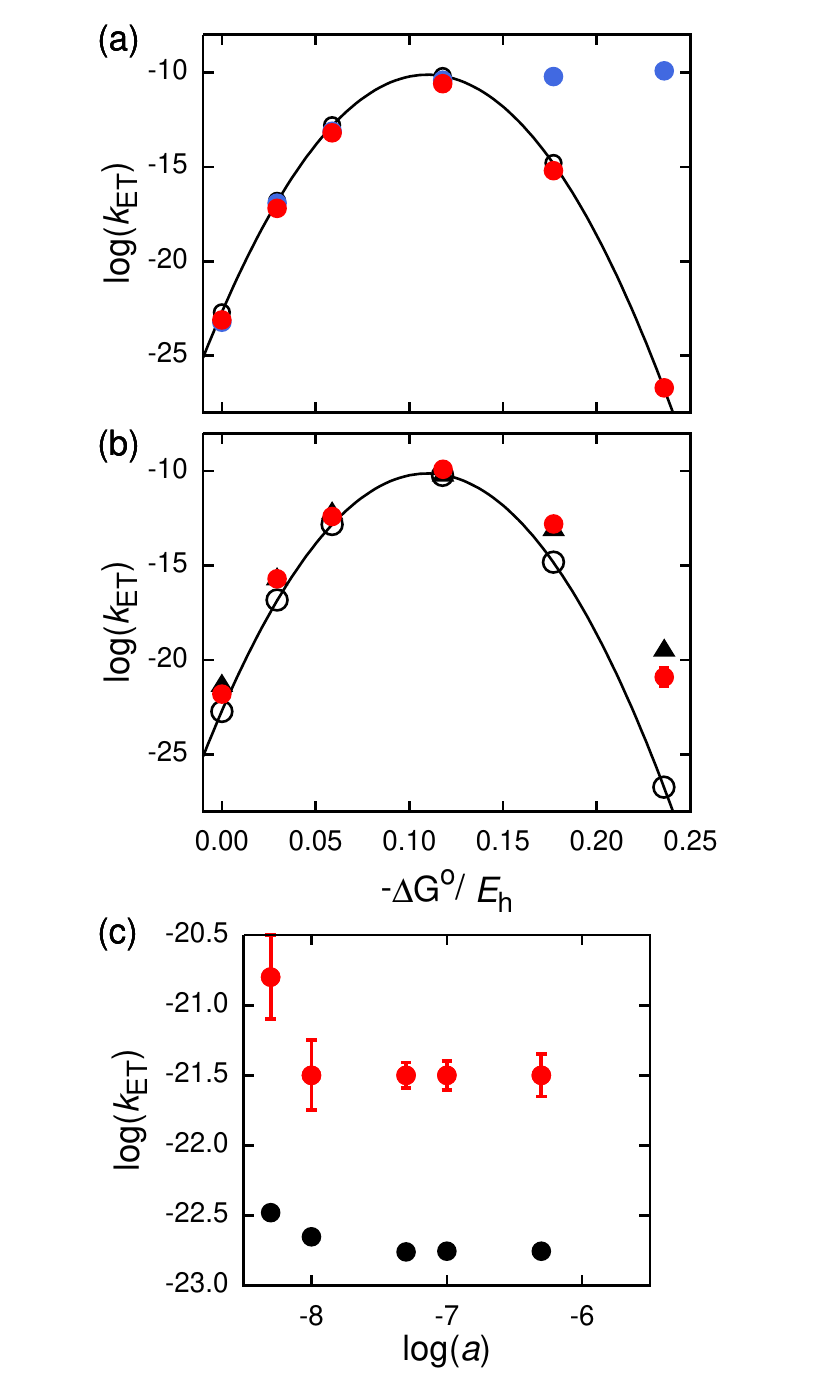}
\caption{(a) 
ET reaction rate coefficients  for System B with  a classical description of the solvent coordinate, obtained as a function of ET driving force using KC-RPMD (red), classical MT (Eq. \ref{eq:knad}, black open circles), and position-representation RPMD (Ref. \onlinecite{Men11}, blue).
(b) The corresponding results for 
System B with  a quantized description of the solvent coordinate, obtained 
using KC-RPMD (red)
and the golden-rule expression in Eq. \ref{eq:qmt2} (black triangles).  Results obtained using classical MT are also included for comparison (black open circles).
(c) The convergence of the KC-RPMD reaction rate for symmetric ET with respect to the strength of kinetic constraint, $a$, including both classical (black) and quantized (red) descriptions of the solvent.
}
\label{fig:cet}
\end{figure}

Figure \ref{fig:cet}(a) presents thermal reaction rates for this system in the weak-coupling regime ($\beta K\approx7\times10^{-4}$)  and for a broad range of the thermodynamic driving force, obtained using KC-RPMD (red), position-representation RPMD (blue), and the non-adiabatic MT relation in Eq. \ref{eq:knad}.  For this set of results, the solvent coordinate is 
treated classically, such that the classical MT relation provides the appropriate reference result. The position-representation RPMD results in this figure are reproduced from Ref. \onlinecite{Men11}. Comparison of the MT results and the position-representation RPMD results in the figure
 reiterate the observations from Ref. \onlinecite{Men11}; this previous implementation of the RPMD method provides an accurate description of the ET rate throughout the normal and activationless regimes of the driving force, but the breakdown of the instanton tunneling rate for strongly asymmetric double-well systems leads to the absence of the rate turnover in the inverted regime. Correction of this breakdown via introduction of the kinetic constraint in the KC-RPMD method (red) leads to quantitative agreement with the reference results across the full range of  driving forces.  
Fig. \ref{fig:cet}(a) clearly demonstrates that, in addition to enabling the use of many-electron wavefunctions in the diabatic representation, the KC-RPMD method successfully avoids the most dramatic known failure of the position-representation RPMD method.

Figure \ref{fig:cet}(b) presents numerical results for System B that include 
quantization of the solvent coordinate. 
The KC-RPMD results are plotted in red, and the results for MT with the classical solvent are re-plotted for reference.  Also included are the golden-rule ET rates from Eq. \ref{eq:knad}, which explicitly include the quantization of the solvent coordinate. Just as KC-RPMD quantitatively reproduced the  MT relation in the limit of classical nuclei (Fig. \ref{fig:cet}(a)), Fig. \ref{fig:cet}(b) demonstrates that KC-RPMD reproduces the effects of nuclear quantization on the ET reaction rate throughout the full range of driving forces.  In particular, nuclear quantization enhances the KC-RPMD rate in the normal regime far less than in the inverted regime, as is consistent with Eq. \ref{eq:qmt}. 

Figure \ref{fig:cet}(c) presents convergence tests for the symmetric ET reaction rate with ($\beta K\approx7\times10^{-4}$), including both classical (black) and quantized (red) descriptions of the solvent.  Specifically, we plot the KC-RPMD rate as a function of the strength of the kinetic restraint, $a$.  In both cases, it is seen that for small values of $a$, the rate varies with $a$ since the kinetic constraint is not fully enforced.  However, for sufficiently large values of $a$, the kinetic constraint is enforced and the rate converges with respect to this parameter. 
Similar results are obtained for the cases with non-zero driving force. 

 \begin{figure}[h]
\includegraphics[scale=1.0]{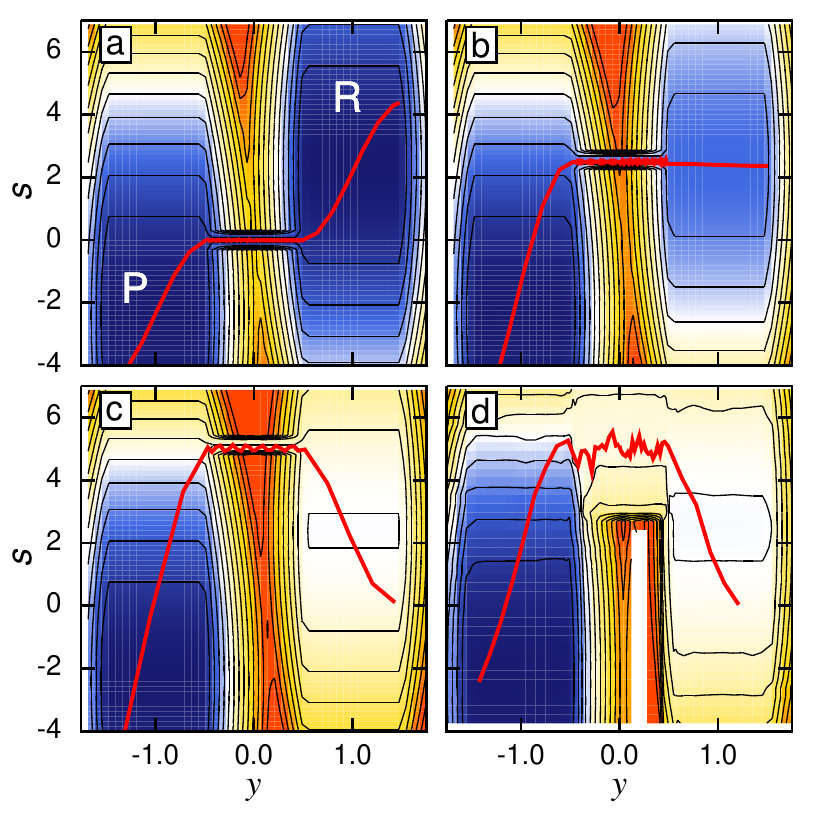}
\caption{(a)-(c)Representative trajectories (red) from the ensemble of reactive KC-RPMD trajectories for the (a) symmetric, (b) activationless, and (c) inverted  regimes of ET, obtained using the classical description of the solvent coordinate. The trajectories are projected onto the plane of the solvent coordinate $s$ and the auxiliary variable $y$. The trajectories overlay the FE surface $F(s,y)$, with contour lines indicating increments of  $0.0475$ $E_\mathrm{h}$ ($50\ k_\mathrm{B} T$).  The ET reactant and product basins are indicated using ``R" and ``P," respectively. (d) The corresponding results for the inverted regime, obtained using the quantized description of the solvent coordinate. To more clearly illustrate the effect of solvent quantization, the trajectories and FE profile are plotted as a function of the solvent ring-polymer bead position, $s^{(\alpha)}$, rather than the centroid position.}
\label{fig:qet_traj} 
\end{figure}


Figures \ref{fig:qet_traj}(a)-(c) present representative reactive KC-RPMD trajectories for System B in the symmetric ($\epsilon=0$), activationless ($\epsilon=0.1178$), and inverted ($\epsilon=0.236$) regimes for ET. 
The solvent is treated classically, and the illustrative trajectories overlay the 2D FE profile $F(s,y)$.
In each case, the 
KC-RPMD trajectories exhibit the reaction mechanism that is anticipated in MT, with distinct components of the trajectories undergoing \emph{(i)} solvent reorganization to configurations for which the electronic diabatic states are nearly degenerate, \emph{(ii)} reactive tunneling of the electron between the redox sites at solvent configurations for which the electronic diabatic states are nearly degenerate, and \emph{(iii)} solvent relaxation in the product basin following reactive tunneling.  As was emphasized in Ref. \onlinecite{Men11}, these features of MT emerge clearly for position-representation RPMD in the normal and activationless regimes, but they do not correctly appear in the inverted regime. By penalizing ring-polymer configurations that lead to the overestimation of reactive tunneling via the kinetic constraint, the KC-RPMD method correctly predicts the solvent-reorganization reaction mechanism for all regimes of the ET driving force. 


Figure \ref{fig:qet_traj}(d) reproduces the results for the inverted regime using the quantized description for the solvent coordinate. 
As for the results obtained with classical solvent (Fig. \ref{fig:qet_traj}(c)), the reactive trajectory exhibits the solvent-reorganization reaction mechanism for the inverted regime.
However, comparison of Figs. \ref{fig:qet_traj}(c) and \ref{fig:qet_traj}(d) reveals in the quantized description for the solvent, widening of the transition channel significantly reduces the degree to which solvent reorganization is needed for reactive tunneling.
By allowing for a degree of ``corner-cutting" in the solvent coordinate, this quantum effect 
gives rise to the significant weakening of the turnover in the ET reaction rate in the inverted regime that is observed in Fig. \ref{fig:cet}(b).

\begin{figure}[h]
\includegraphics{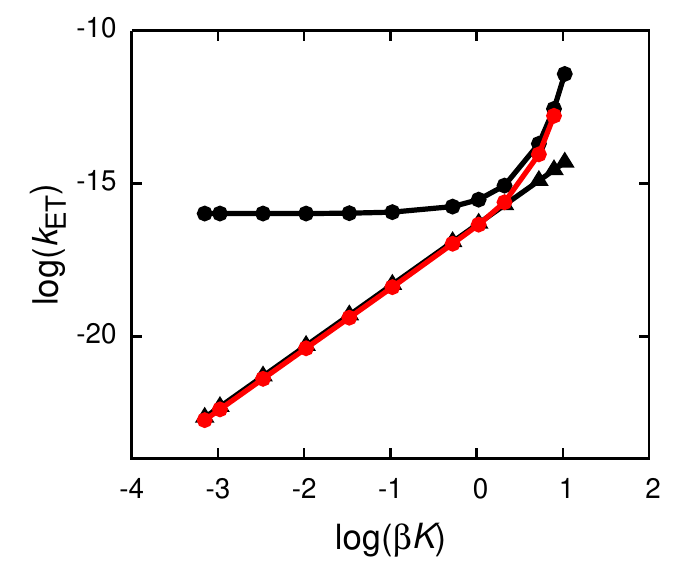}
\caption{
ET reaction rate coefficients  for System B with  a classical description of the solvent coordinate, obtained as a function of the non-adiabatic coupling using KC-RPMD (red), the non-adiabatic rate expression in Eq. \ref{eq:knad} (black triangles), and the adiabatic rate expression in Eq. \ref{eq:kad} (black circles).}
\label{fig:adnad} 
\end{figure}

Finally, Figure \ref {fig:adnad}  presents   rate coefficients for System B obtained over a range of values for the non-adiabatic coupling $K$ that span from the weak-coupling to the strong-coupling regimes.  In all cases, $\epsilon=0$, and the solvent degree of freedom is treated classically. 
For comparison with the KC-RPMD reaction rates (red), reference results are included  from rate expressions that are derived in the non-adiabatic regime (Eq. \ref{eq:knad}, black triangles) and in the adiabatic regime (Eq. \ref{eq:kad}, black circles). 
Although the KC-RPMD method makes no  \emph{a priori} assumption about the coupling regime for the reaction, it is seen that the method quantitatively reproduces the reference results in the appropriate regimes, and the KC-RPMD method correctly transitions from the non-adiabatic result to the adiabatic result in the regime of intermediate coupling $(\textrm{log}(\beta K)\approx0)$.

\section {Concluding Remarks}
\label{sec:conclusion}
The development of accurate and robust methods for describing non-adiabatic chemistries in complex, condensed-phase systems is a central methodological challenge for the field of molecular simulation. 
In this work, we present an extension of  RPMD  that is well suited to addressing this challenge for broad classes of donor-acceptor chemistries.
The KC-RPMD method is a path-integral-based method that provides continuous equations of motion to model the non-adiabatic molecular dynamics of systems that are quantized with respect to both electronic and nuclear degrees of freedom.
The method generates trajectories that rigorously preserve a well-defined equilibrium distribution, such that  KC-RPMD  exhibits the appealing features of the previously formulated position-representation RPMD method, including  detailed balance, time-reversal symmetry, and  invariance of reaction rate calculations to the choice of  dividing surface.  
The distribution that is preserved in  KC-RPMD  is modified from the exact quantum Boltzmann distribution by introducing a kinetic constraint to penalize 
ring-polymer configurations that make a small contribution to the thermal ensemble but that lead to the overestimation of deep-tunneling rates across asymmetric barriers.
 KC-RPMD yields very encouraging results for a range of condensed-phase charge-transfer chemistries, as is demonstrated using model systems that investigate the performance of the method for  adiabatic vs. non-adiabatic reactions, classical vs. quantized nuclei, and  normal vs. inverted ET.
We emphasize that KC-RPMD is computationally efficient (with force-evaluations that scale linearly with the number of ring-polymer beads), relatively easy to perform (as it simply involves the integration of continuous classical-like equations of motion), naturally interfaced with  electronic structure packages (as the electronic states correspond to general, many-electron wavefunctions in the diabatic representation),  and free of uncontrolled parameters.
Furthermore, the method enables the immediate and straightforward utilization of the full toolkit of classical molecular dynamics simulation, including rare-event sampling methods, and it is robustly scalable to large, complex systems.
We expect that it will prove useful for the simulation of charge-transfer and non-adiabatic chemistries in a range of future applications.

\section{Acknowledgments}

This work was supported by the National Science Foundation (NSF) CAREER Award under Grant No. CHE-1057112, the (U.S.) Department of Energy (DOE) under Grant No. DE-SC0006598, and the Office of Naval Research (ONR)  under Grant No. N00014-10-1-0884. Additionally, T.F.M. acknowledges support from a Camille and Henry Dreyfus Foundation New Faculty Award and an Alfred P. Sloan Foundation Research Fellowship. Computing resources were provided by the National Energy Research Scientific Computing Center (NERSC) (DE-AC02-05CH11231) and the Oak Ridge Leadership Computing Facility (OLCF) (DE-AC05-00OR22725).  The authors sincerely thank David Chandler, David Manolopoulos, William Miller, and Nandini Ananth for helpful conversations.

%
%




\appendix{


\section{Derivation of the penalty function}
\label{app:c_prefs}

In this appendix, we derive the specific form of the penalty function, $g$, that appears in Eq. \ref{eq:g_func}.  
The penalty function  enforces the kinetic constraint by restraining the formation of kinked configurations of the ring polymer to the region of the crossing of the diabatic surfaces (thereby excluding ring-polymer configurations that have low thermodynamic weight in the equilibrium ensemble but which contribute substantially to the incorrect instanton TST estimate for the rate). This is accomplished by a Gaussian function that is centered at the intersection of diabatic surfaces, with the energy scale set by the non-adiabatic coupling, $K$, such that
\begin{equation}
g(\{i^{(\alpha)}\},\allq)\!=\!\left\{
\begin{array}{c l}
1,&i^{(\alpha)}=0\ \mbox{for all $\alpha$,}\\
1,&i^{(\alpha)}=1\ \mbox{for all $\alpha$,}\\
C e^{ - aw^2(\bar \rvec) }, &\mbox{otherwise},
\end{array}\right.
\end{equation}
where $C$ is a multiplicative prefactor, and $w$ is defined in the main text (after Eq. \ref{eq:g_func}).
We choose a form for the penalty function in which the intersection of the diabatic surfaces is defined in terms of the centroid of the ring polymer, which is convenient and  has a natural classical limit; however, other sensible choices of the penalty function are possible.

To avoid biasing the rate of reactive tunneling at the  nuclear configurations for which the diabats cross, we require that the FE of kink-pair formation is unchanged by the kinetic constraint at these nuclear configurations, and we derive the expression for $C$ based on this condition.  
Specifically, we consider the FE 
cost of going from unkinked configurations of the ring polymer in the reactant basin to kinked configurations at the crossing of the diabatic surfaces, and we equate this to the FE cost of kink-pair formation at the intersection of the diabats in the unmodified distribution. 

For simplicity, we first present the detailed derivation for a 1D redox system with constant coupling, $K$, in the classical limit for the nuclear coordinate. We then outline the analogous derivations for a 1D redox system with quantized nuclei and for a general multi-dimensional system. 

\subsection{1D redox system with constant $K$ and classical nuclei}
\label{sec:1d_cm}

 For a 1D system with classical nuclei, the kinetically constrained ring-polymer distribution (Eq. \ref{eq:rho_kc}) has the form
\begin{equation}
\rho_n^\mathrm{KC}(x,y)\!=\! \Omega \!\sum_{\{i^{\alpha}\}}    g(\{i^{(\alpha)}\}, x) e^{-\beta V_{\textrm{r}}(y, \{i^{(\alpha)} \})} \Gamma(\{i^{(\alpha)}\},x), 
\end{equation}
where $\Gamma (\{i^{(\alpha)}\}, x)= \prod_{\alpha=1}^n \! M_{i^{(\alpha)},i^{(\alpha+1)}}\!(x)$, and the penalty function in this case takes the form
 \begin{equation}
g(\{i^{(\alpha)}\},x)\!=\!\left\{
\begin{array}{c l}
1,&i^{(\alpha)}=0\ \mbox{for all $\alpha$,}\\
1,&i^{(\alpha)}=1\ \mbox{for all $\alpha$,}\\
Ce^{ - aw^2( x ) }, &\mbox{otherwise}.
\end{array}\right.
\end{equation} 

In the kinetically constrained distribution, the FE cost of 
 going from unkinked configurations of the ring polymer in the reactant basin to kinked configurations at the crossing of the diabatic surfaces is
 $F^{\ddagger}=-\frac{1}{\beta}\ln P^{\textrm{KC}}(y=y^\ddagger)$, where 
\begin{align}
\label{eq:py1}
P^{\textrm{KC}}(y=y^\ddagger) = Z_0^{-1} e^{-\beta \Delta F(y^\ddagger)}, 
\end{align}
\begin{equation}
Z_0 = \int_{-\infty}^{y^\ddagger} dy'e^{-\beta \Delta F(y')},
\end{equation}
\begin{equation}
e^{-\beta \Delta F(y)} = 
\int d\allq\ e^{-\beta V_\mathrm{eff}(\allq,y)},
\end{equation}
and $y^\ddagger=0$. 

For kinked ring-polymer configurations (i.e., $y~=~y^\ddagger$), the numerator on the right-hand side (RHS) of Eq. \ref{eq:py1} simplifies to
\begin{align*}
\label{eq:num_cl}
e&^{-\beta \Delta F(y^\ddagger)} = \int dx   \ e^{-\beta V_\mathrm{eff}(x,y^\ddagger)} \numberthis\\
& \!= \!  C \int dx\sum_{\{i^{\alpha}\}} \mathcal{P}_{k} (\{i^{(\alpha)}\}) e^{-aw^2(x)} \Gamma (\{i^{(\alpha)}\}, x) \\
&\!= \!  C \int dx\;e^{-a (w(x))^2} \sum_{k=1}^{n/2} \frac{{(\beta K)}^{2k}} {\phi_n(k)} \frac{ e^{-\beta V_0(x)} - e^{-\beta V_1(x)}}{\beta (V_1(x)-V_0(x))} ,
\end{align*}
where $\phi_n (k)=\left(\frac{2}{n^{2k}} {n \choose 2k }\right)^{-1}$, and $\mathcal{P}_{k} (\{i^{(\alpha)}\})$ is unity for configurations characterized by $k$ kink-pairs and $0$ otherwise.   
The last equality in Eq. \ref{eq:num_cl} is obtained by evaluating the 
sum over ring-polymer configurations in the limit of large $n$.\cite{Stu68}  

A consequence of the penalty function is that only nuclear configurations in the vicinity of the intersection of the diabatic surfaces contribute to the integral over $x$. Therefore, for sufficiently large values of $a$, the penalty function tends to a Dirac $\delta$-function,
\begin{equation}
\label{eq:deltaid}
\lim_{a \to\infty}e^{-a (w(x))^2}  = \delta(w(x)) \sqrt{\frac{\pi }{a}}.
\end{equation}
Using this identity  and performing the integral over $x$, Eq. {\ref{eq:num_cl}} becomes 
\begin{align*}
\label{eq:num_cl2}
e&^{-\beta \Delta F(y^\ddagger)} \\
&= C  \sqrt{\frac{\pi }{a}}   \sum_{k=1}^{n/2} \frac{{(\beta K)}^{2k}} {\phi_n (k)} \!\! \int \!\!dx\  \delta(w(x))\frac{ e^{-\beta V_0(x)} - e^{-\beta V_1(x)}}{\beta (V_1(x)-V_0(x))}     \\
 &=C  \sqrt{\frac{\pi }{a}}  \sum_{k=1}^{n/2} \frac{{(\beta K)}^{2k}} {\phi_n (k)} \ e^{-\beta V_{0} (x^\ddagger)} \left|w'(x^\ddagger)\right|^{-1}, \numberthis
\end{align*}
where $x^\ddagger$ denotes the point of the intersection of the diabatic surfaces (the solution of $w(x)=0$), and  the prime denotes differentiation with respect to the nuclear coordinate. 

We now consider the denominator $Z_0$ in Eq. \ref{eq:py1}, which 
is dominated by the statistical weight of unkinked configurations. For these configurations, the penalty function makes no contribution, such that 
\begin{align*}
\label{eq:refyo}
Z_0 &= \int_{-\infty}^{y^\ddagger} dy \int dx e^{-\beta V_\mathrm{eff}(x,y)} \numberthis \\
& =  \int_{-\infty}^{y^\ddagger} dy \int dx\ f(y,-1) \ \Gamma(\{0\},x), 
\end{align*}
where we have used the definition of $f(y, \theta(\{i^\alpha\}))$ from Eq. \ref{eq:vrestraint}, and $\{0\}$ denotes ring-polymer configurations which have $i^{(\alpha)}=0$ for all $\alpha$.
Inserting the definition of $\Gamma(\{0\},x)$ into the RHS of Eq. \ref{eq:refyo} yields 
\begin{align*}
\label{eq:refyo2}
Z_0 &= \int_{-\infty}^{y^\ddagger} dy \int dx\ f(y,-1) \ e^{-\beta V_{0} (x)}\numberthis \\
& =   \int dx \ e^{-\beta V_{0} (x)}.
\end{align*}

Combining the results of Eqs. \ref{eq:py1}, \ref{eq:num_cl2}, and \ref{eq:refyo2}, we obtain
the probability of forming kinked ring-polymer configurations at the crossing of the diabatic surfaces in the kinetically constrained distribution, 
\begin{align*}
\label{eq:py2}
P^{\textrm{KC}}(y=y^\ddagger) = &\frac{e^{-\beta V_{0} (x^\ddagger)}}{\int dx \ e^{-\beta V_{0} (x)}} \times \numberthis \\
&\quad \frac{C}{|w'(x^\ddagger)|} \sqrt{\frac{\pi }{a}} \  \sum_{k=1}^{n/2} \frac{{(\beta K)}^{2k}} {\phi_n (k)}
\end{align*}
Here, the first term on the RHS corresponds to the  FE cost of reorganizing the nuclear coordinates to configurations for which the diabatic surfaces are degenerate, and the second term corresponds to the FE cost for ring-polymer kink-pair formation at the reorganized nuclear configurations and in  the presence of the penalty function. 
The analog of Eq. \ref{eq:py2} for the ring-polymer distribution without the kinetic constraint (i.e., in the absence of the penalty function) is 
\begin{equation}
\label{eq:py3}
P(y=y^\ddagger) = \frac{e^{-\beta V_{0} (x^\ddagger)}}{\int dx \ e^{-\beta V_{0} (x)}}  \sum_{k=1}^{n/2}\frac{{(\beta K )^{2k}}}{ \phi_n (k)} . 
\end{equation}
Finally, enforcing the condition that the probabilities in Eqs. \ref{eq:py2} and \ref{eq:py3} are identical yields the final expression for the multiplicative prefactor
in a 1D redox system with constant $K$ and classical nuclei,
\begin{equation}
\label{eq:c_1d}
C=\sqrt{\frac{a}{\pi}} |w'(x^\ddagger)|.
\end{equation}

\subsection{1D redox system  with constant $K$ and quantized nuclei}
\label{sec:1d_qm}

We now repeat the derivation of $C$ for the case of a 1D redox system with constant $K$ and quantized nuclei. 
In this case, the steps 
outlined in Eqs. \ref{eq:num_cl}-\ref{eq:num_cl2} yield 
\begin{align*}
\label{eq:num_1dqm}
  e^{-\beta \Delta F(y^\ddagger)} \numberthis  
  & \!=\! C \sqrt{\frac{\pi }{a}}  \int \!\!d\xvec\  \delta(w(\bar x)) e^{-\beta U_{\mathrm{int}}(\xvec)} \Phi(\xvec)\\
 & \!=\!C \sqrt{\frac{\pi }{a}}  \int \!\!d\xvec\  \delta(\bar{x}-x^\ddagger) e^{-\beta U_{\mathrm{int}}(\xvec)}\frac{\Phi(\xvec)}{|w'(\bar{x})|}.
\end{align*}
where $\xvec$ denotes the vector of ring-polymer position coordinates $\{x^{(\alpha)}\}$, $\bar{x}$ is the centroid of the ring polymer, and 
\begin{equation}
\label{eq:phi}
\Phi (\xvec) \!=\!\left ( \! \mathrm{Tr} \! \prod_{\alpha=1}^n \mathbf{M}(x^{(\alpha)}) \!- \! \prod_{\alpha=1}^n \! M_{0,0}(x^{(\alpha)})\!- \!  \prod_{\alpha=1}^n \!  M_{1,1}(x^{(\alpha)} ) \!\! \right ).
\end{equation}
As before, $Z_0$ in 
Eq. \ref{eq:py1} is unaffected by the penalty function, and it simplifies in this case to 
\begin{equation}
\label{eq:denom_1dqm}
Z_0 = 
\int d\xvec e^{-\beta U_{\mathrm{int}}(\xvec)}  \prod_{\alpha=1}^n \! M_{0,0}(x^{(\alpha)}).
\end{equation}


Combining the results of Eqs. \ref{eq:py1}, \ref{eq:num_1dqm}, and \ref{eq:denom_1dqm}, we obtain
the probability of forming kinked ring-polymer configurations at the crossing of the diabatic surfaces in the kinetically constrained distribution, 
\begin{align*}
\label{eq:py_qm2}
P^{\textrm{KC}}(y&=y^\ddagger) =  \frac{C}{Z_0} \sqrt{\frac{\pi }{a}}  \int \!\!d\xvec\  \delta(\bar{x}-x^\ddagger) e^{-\beta U_{\mathrm{int}}(\xvec)}\frac{\Phi(\xvec)}{|w'(\bar{x})|}. \numberthis
\end{align*}
The analog of Eq. \ref{eq:py_qm2} for the ring-polymer distribution without the kinetic constraint is 
\begin{align*}
\label{eq:ref_qm}
P(y=y^\ddagger) &=
  Z_0^{-1} \int \!\!d\xvec\  \delta(\bar{x}-x^\ddagger) e^{-\beta U_{\mathrm{int}}(\xvec)} \Phi(\xvec) \numberthis\\
  &= Z_0^{-1} \int \!\!d\xvec\  \delta(w(\bar{x})) |w'(\bar{x})| e^{-\beta U_{\mathrm{int}}(\xvec)} \Phi(\xvec). 
\end{align*}
Finally, enforcing the condition that the probabilities in Eqs. \ref{eq:py_qm2} and \ref{eq:ref_qm} are identical yields the final expression for the multiplicative prefactor
in a 1D redox system with constant $K$ and quantized nuclei,
\begin{equation}
\label{eq:c_1dqm}
C=\sqrt{\frac{a }{\pi}}  \frac{\int \!\!d\xvec\  \delta(w(\bar{x})) |w'(\bar{x})| e^{-\beta U_{\mathrm{int}}(\xvec)}
\Phi(\xvec)}{ \int \!\!d\xvec\  \delta(w(\bar{x})) e^{-\beta U_{\mathrm{int}}(\xvec)} \Phi(\xvec)}.
\end{equation}
Equation \ref{eq:c_1dqm} has the form of a constrained ensemble average, 
which can be evaluated using standard methods. 

If the ring-polymer nuclear coordinates are approximated by the centroid position, $\Phi(\xvec)$ can be further simplified as follows, 
\begin{equation}
\label{eq:phi_approx}
\Phi(\xvec)=    \frac{ e^{-\beta V_0(\bar{x})} - e^{-\beta V_1(\bar{x})}}{\beta (V_1(\bar{x})-V_0(\bar{x}))} \sum_{k=1}^{n/2}\frac{(\beta K )^{2k}} {\phi_n(k)}. 
\end{equation}
Inserting Eq. \ref{eq:phi_approx}  into 
Eq. \ref{eq:c_1dqm} 
yields the final result for the multiplicative prefactor in a 1D redox system with quantized nuclei,
\begin{equation}
\label{Cv2}
C=\sqrt{\frac{a}{\pi}} |w'(x^\ddagger)|.
\end{equation}
Note that this result is identical to that obtained for a system with classical nuclei in Eq. \ref{eq:c_1d}. 
Furthermore, note that Eqs. \ref{eq:c_1dqm} and \ref{Cv2} are identical 
in the limit of classical nuclei or for a quantized system with constant coupling and harmonic diabatic potentials.

\subsection{Multi-dimensional redox system with position-dependent $K (\rvec)$}
\label{sec:c_md}

For the case of a general multi-dimensional system with classical nuclei and $\rvec$-dependent non-adiabatic coupling $K (\rvec)$, the previously outlined derivation yields
\begin{equation}
\label{eq:c_md_0}
C\!=\!\sqrt{\frac{a}{\pi}} \langle | \nabla w(\rvec)| \rangle_\Sigma,
\end{equation} 
where  the brackets denote a constrained ensemble average constrained to at the hypersurface $w(\rvec)=0$,
\begin{equation}
\label{eq:md_constraint}
\langle \dots \rangle_\Sigma =  \frac{ \sum_{k=1}^{n/2} \frac{ {(\beta)}^{2k}} {\phi_n(k)} \int\! d\rvec \ \delta(w(\rvec)) (\dots) |K(\rvec)|^{2k}e^{-\beta  V_{0}(\rvec)}} { \sum_{k=1}^{n/2} \frac{ {(\beta)}^{2k}} {\phi_n(k)} \int\! d\rvec \ \delta(w(\rvec)) |K(\rvec)|^{2k}e^{-\beta  V_{0}(\rvec)}}.
\end{equation}
This expression can be further simplified if it is assumed that terms associated with more than one kink-pair ($k=1$) can be neglected in both the numerator and denominator.
The resulting expression is
\begin{equation}
\label{eq:c_md}
C=\sqrt{\frac{a}{\pi}} \langle |\nabla w({\rvec})| \rangle_\textrm{c},
\end{equation}
where the brackets denote an ensemble average constrained to the intersection of the diabatic surfaces, as described in Eq. \ref{eq:const_avg}. 
We note that Eqs. \ref{eq:c_md_0} and \ref{eq:c_md} are identical for the case of constant non-adiabatic coupling,  $K$, and Eq. \ref{eq:c_md}
 reduces to Eq. \ref{eq:c_1d} for the case of a 1D redox system. 
 
Finally, following the approach described in Section \ref{sec:1d_qm}, the multiplicative prefactor for the case of a general multi-dimensional system with quantized nuclei and $\rvec$-dependent non-adiabatic coupling is derived to be 
\begin{equation}
C=\sqrt{\frac{a }{\pi}}  \frac{\int \!\!d\allq\  \delta(w(\bar{\rvec})) |\nabla w(\bar{\rvec})| e^{-\beta U_{\mathrm{int}}(\allq)}
\Phi(\allq)}{ \int \!\!d\allq\  \delta(w(\bar{\rvec})) e^{-\beta U_{\mathrm{int}}(\allq)} \Phi(\allq)}.
\end{equation}
Employing the approximation for $\Phi(\allq)$ described in Eq. \ref{eq:phi_approx} and again truncating the sums in the numerator and denominator at terms associated with a single kink-pair, we arrive at the same result that was obtained for a system with classical nuclei in Eq. \ref{eq:c_md}, 
\begin{equation}
C=\sqrt{\frac{a}{\pi}}\langle |\nabla w(\rvec)| \rangle_\textrm{c}.
\end{equation}
This  expression for the multiplicative prefactor appears in the main text in Eq. \ref{eq:g_func}.

\section{KC-RPMD forces and the Bell algorithm}

\label{app:bellalg}
In this appendix, we illustrate the terms that arise in the calculation of forces associated with the KC-RPMD effective potential ($V_{\mathrm{eff}}^{\mathrm{KC}}(\allq,y)$ in Eq. \ref{eq:veff}), and we review a computational algorithm\cite{bellthesis} that enables the evaluation of these forces with a cost 
that scales linearly with 
the number of ring-polymer beads.

Without approximation, the KC-RPMD effective potential can be factorized to obtain 
\begin{align*}
&V_{\mathrm{eff}}^{\mathrm{KC}}(\allq,y)=  U_{\mathrm{int}}(\allq)  \\
& \quad\quad\quad -\frac{1}{\beta}\ln \bigg[ f(y,0) \left(\frac{a}{\pi}\right)^{\frac{1}{2}}\eta e^{ - aw^2(\bar\rvec) } \times \numberthis \\  
& \; \; \left ( \! \mathrm{Tr} \! \prod_{\alpha=1}^n \mathbf{M}(\rvec^{(\alpha)}) \!- \! \prod_{\alpha=1}^n \! M_{0,0}(\rvec^{(\alpha)})\!- \!  \prod_{\alpha=1}^n \!  M_{1,1}(\rvec^{(\alpha)} ) \!\! \right )     \\ 
 &\quad \left.+f(y,-1)\!\prod_{\alpha=1}^n\! \!M_{0,0} (\rvec^{(\alpha)}) + f(y,1)\! \prod_{\alpha=1}^n\! M_{1,1}(\rvec^{(\alpha)})   \right] .
\end{align*}
Differentiation of this term with respect to a given nuclear coordinate $\xi^{(\alpha)}$ 
leads to terms of the form 
\begin{align*}
\label{eq:grad}
\frac{\partial }{\partial \xi^{(\alpha)}}\!\! \left[ \ln \left (\mathrm{Tr} \prod_{\alpha =1}^{n} \right.\right.&\left.\left.\!\! \mathbf{M}(\mathbf{R}^{(\alpha)}) \right)\! \right] \numberthis \\
& = \! \frac{  \mathrm{Tr} \left[ \mathbf{F}_{\alpha-1} \mathbf{{D}}^\xi_\alpha\mathbf{G}_{\alpha+1}\right]}{ \mathrm{Tr}\left[ \prod_{\alpha}^{n} \mathbf{M}(\mathbf{R}^{(\alpha)})\right]}, 
\end{align*}
where 
\begin{equation}
\mathbf{F}_{\alpha-1} = \mathbf{M}(\mathbf{R}^{(1)}) \mathbf{M}(\mathbf{R}^{(2)}) \dots \mathbf{M}(\mathbf{R}^{(\alpha-1)}),
\end{equation}
\begin{equation}
\mathbf{G}_{\alpha+1} = \mathbf{M}(\mathbf{R}^{(\alpha+1)}) \mathbf{M}(\mathbf{R}^{(\alpha+2)}) \dots \mathbf{M}(\mathbf{R}^{(n)}),
\end{equation}
and
\begin{equation}
\begin{split}
\label{Dx}
\mathbf{D}^\xi_\alpha \!=\!\frac{\partial}{\partial \xi^{(\alpha)}} \mathbf{M}(\mathbf{R}^{(\alpha)}).
\end{split}
\end{equation} 
Using the cyclic property of the trace, the numerator of Eq. \ref{eq:grad} can be expressed 
\begin{equation}
\begin{split}
\mathrm{Tr} \left[ \mathbf{F}_{\alpha-1} \mathbf{D}^\xi_\alpha \mathbf{G}_{\alpha+1}\right]=
 \mathrm{Tr} \left[ \mathbf{{D}}^\xi_\alpha \mathbf{H}_\alpha\right],
\end{split}
\end{equation}
where $\mathbf{H}_\alpha$ is the `hole' matrix that is given by 
\begin{align*}
\label{eq:holematrix}
\mathbf{H}_\alpha \!&=\mathbf{G}_{\alpha+1} \mathbf{F}_{\alpha-1} \numberthis\\
&=\!\mathbf{M}(\mathbf{R}^{(\alpha+1)}) \dots \mathbf{M}(\mathbf{R}^{(n)}) \mathbf{M}(\mathbf{R}^{(1)}) \dots \mathbf{M}(\mathbf{R}^{(\alpha-1)}).
\end{align*}
Since the matrices $\mathbf{M}(\mathbf{R}^{(\alpha)})$ do not generally commute, a naive algorithm would individually determine the hole matrix for each ring-polymer bead, 
at a combined cost of that entails $\mathcal{O}(n^2)$  matrix multiplications. 
Using the algorithm outlined below, however, only $\mathcal{O}(n)$ matrix multiplications are required. 

\subsection{The Bell algorithm}  
The gradients of  $V_{\mathrm{eff}}^{\mathrm{KC}}(\allq,y)$ can be efficiently evaluated by taking advantage of 
the appearance of common terms in the hole matrices for different ring-polymer beads.\cite{bellthesis}  
By calculating and storing portions of these matrices, the overall time for the calculation is greatly reduced. The algorithm is clearly outlined in Ref.~\onlinecite{helethesis} and proceeds as follows. 
\begin{enumerate}
\item Set $\mathbf{F}_1 = \mathbf{M}(\mathbf{R}^{(1)})$ and compute $\mathbf{F}_{\alpha}$ for $\alpha = 2,\dots, n-1$ recursively,  noting that $\mathbf{F}_{\alpha} = \mathbf{F}_{\alpha-1} \mathbf{M}(\mathbf{R}^{(\alpha)})$. This step requires  $n-2$ matrix multiplications. 
\item Set $\mathbf{G}_n = \mathbf{M}(\mathbf{R}^{(n)})$ and compute $\mathbf{G}_{\alpha}$,  $\alpha=n-1,n-2,\dots,2$ recursively,  noting that $\mathbf{G}_{\alpha} = \mathbf{M}(\mathbf{R}^{(\alpha)})\mathbf{G}_{\alpha+1}$. This step requires $n-2$ matrix multiplications. 
\item Compute $\mathbf{H}_\alpha$ for $\alpha = 1,\dots,n$ using Eq. \ref{eq:holematrix}.  This only requires $n-2$ matrix multiplications because $\mathbf{H}_1= \mathbf{G}_2$ and $\mathbf{H}_n= \mathbf{F}_{n-1}$. 
\end{enumerate}
With this algorithm, all the  $\mathbf{H}_\alpha$ matrices required for evaluation of the gradients of  $V_{\mathrm{eff}}^{\mathrm{KC}}(\allq,y)$ are constructed in $3n-6$ matrix multiplications.



\section{Derivation of the mass of the auxiliary variable}
\label{app:mass}

In this appendix, we derive 
the mass of auxiliary variable, $m_y$, which is chosen such that the
KC-RPMD TST recovers the Landau-Zener (LZ) TST \cite{Lan32,Zen32} in the limit of weak non-adiabatic coupling.
 We first describe the case of a 1D redox system with classical nuclei and 
 constant non-adiabatic coupling, before outlining the 
 general case of  multi-dimensional system with position-dependent  non-adiabatic coupling and quantized nuclei.
  
\subsection{1D redox system  with constant $K$ and classical nuclei}

The LZ TST rate for a non-adiabatic process in 1D is given by\cite{Nitzanbook}
\begin{equation}
\label{rate1}
k_{\mathrm{TST}}^{\mathrm{LZ}} =\int_0^{\infty} d\xdot \xdot P(\xdot,x^\ddagger) P_{0 \to 1}(\xdot), 
\end{equation}
where $P(\xdot,x^\ddagger)$ denotes the probability of reaching the diabatic crossing $x=x^\ddagger$ with velocity $\xdot$ and $P_{0 \to 1}(\xdot)$ indicates the non-adiabatic transition probability for a given $\xdot$. 
The probability of reaching the diabatic crossing is 
\begin{equation}
\label{probdiab}
P(\xdot,x^\ddagger) = \frac{1}{Q_{R}} \int_{-\infty}^{\infty} dx \delta (x-x^\ddagger)e^{-\beta \left[\frac{1}{2}m\xdot^2 + V_0(x)\right]},
\end{equation}
where $Q_{R}$ is the reactant partition function, which takes the form 
\begin{equation}
Q_{R}=\left (\frac{2 \pi }{\beta m}\right)^{1/2} \int dx e^{-\beta V_{0}(x)}.
\end{equation}
The probability of a non-adiabatic transition 
under the assumption of small, constant coupling $K$ is 
\cite{Lan32, Zen32}
\begin{equation}
\label{1DLZ}
P_{0 \to 1}(\xdot)=\left [  \frac{2 \pi  |K|^2}{\hbar \dot x |V'_0(x)-V'_1(x)|} \right]_{x=x^\ddagger}.
\end{equation}
Inserting Eqs. \ref{probdiab}-\ref{1DLZ} into Eq. \ref{rate1} and evaluating the velocity integral yields the LZ TST rate
\begin{equation}
\label{LZ1D}
k_{\mathrm{TST}}^{\mathrm{LZ}} = \frac{ \pi }{\hbar} \frac{ |K|^2} {|V'_1(x)-V'_0(x)|}_{x=x^\ddagger} \frac{e^{-\beta V_0(x^\ddagger)}}{\int dx e^{-\beta V_{0}(x)}}.
\end{equation}

The KC-RPMD TST rate associated with the $y^\ddagger=0$ dividing surface takes the form 
\begin{equation}
\label{KCRPMDTST}
\begin{split}
k_{\mathrm{TST}}^{\mathrm{KC-RPMD}} &= \sqrt {\frac{ 1}{2 \pi \beta m_y } }  \frac{  e^{-\beta \Delta F(y^\ddagger)} } {\int_{-\infty}^{y^\ddagger} dye^{-\beta \Delta F(y)}},
\end{split}
\end{equation}
which in the low-coupling limit
can be expressed as 
\begin{equation}
\label{KCRPMDTSTfinal}
k_{\mathrm{TST}}^{\mathrm{KC-RPMD}} \!=\! |K|^2 \beta^2 \sqrt {\frac{ 1}{2 \pi \beta m_y } }\! \frac{ e^{-\beta V_{0} (x^\ddagger)} }{\int dx e^{-\beta V_{0} (x)} } .
\end{equation}

Equating the rate expressions 
in Eqs. \ref{LZ1D} and 
\ref{KCRPMDTSTfinal} and solving for the mass of the auxiliary variable yields
\begin{equation}
m_y 
=\frac{\beta^3 \hbar^2}{2 \pi^3} \left|V'_1(x) - V'_0 (x)\right|^2_{x=x^\ddagger}.
\end{equation}

\subsection{Multi-dimensional redox system with position-dependent $K (\rvec)$}

For a general multi-dimensional redox system, the auxiliary-variable mass $m_y$ can be analogously derived. 
In this case, the non-adiabatic coupling $K(\rvec)$ can vary along the seam of crossing of the diabatic surfaces. 
Using the multi-dimensional analogue of the LZ non-adiabatic transition probability,\cite{Sti76} Eq. \ref{rate1} for the general case becomes
\begin{equation}
k_{\mathrm{TST}}^{\mathrm{LZ}}  = \frac{\pi}{\hbar} \frac{\int d\rvec  \delta (\xi(\rvec))   |K(\rvec)|^2  e^{-\beta V_{0}(\rvec)}  }{ \int d\rvec e^{-\beta V_{0}(\rvec)}},
\end{equation}
where $\xi(\rvec)=V_{0}(\rvec)-V_{1}(\rvec)$. 
If we assume that the non-adiabatic coupling is constant in the direction perpendicular to the crossing of the diabatic surfaces, such that
\begin{equation}
\label{eq:delta_assumption}
\left. \nabla \left( K(\rvec)\right) \cdot \nabla \xi(\rvec)\right|_{\xi(\rvec)=0} =0, 
\end {equation}
then this result can be expressed as follows,  
\begin{equation}
\label{multiLZTST}
k_{\mathrm{TST}}^{\mathrm{LZ}}  = \frac{\pi}{\hbar} \frac{\int d\rvec \delta (w(\rvec)) |K(\rvec)|^{-1} |K (\rvec)|^2  e^{-\beta V_{0}(\rvec)}  }{ \int d\rvec e^{-\beta V_{0}(\rvec)}}.
\end{equation}

In analogy to Eq. \ref{KCRPMDTSTfinal}, the KC-RPMD TST rate associated with the $y^\ddagger=0$ dividing surface 
can be expressed 
\begin{align*}
\label{multiRPTST}
k_{\mathrm{TST}}^{\mathrm{KC-RPMD}} =  \sqrt {\frac{ \beta^3}{2 \pi m_y }} &\langle |\nabla w ( \rvec)| \rangle_{\mathrm{c}} \times \numberthis \\
& \frac{ \int d \rvec \; \delta(w(\rvec)) |K(\rvec)|^2 e^{-\beta  V_{0}(\rvec)}  }{\int d\rvec e^{-\beta V_{0}(\rvec)} }. 
\end{align*}

Equating the rate expressions 
in Eqs. \ref{multiLZTST} and 
\ref{multiRPTST}
and solving for $m_y$ yields the final expression for 
a multi-dimensional system with classical nuclei,
\begin{equation}
\label{eq:gen_mass}
m_y=\frac{\beta^3 \hbar^2}{2 \pi^3}\left[  \frac{ \langle |\nabla w ( \rvec)| \rangle_{\mathrm{c}}  }{\langle | K( \rvec )|^{-1} \rangle_{\mathrm{c}} } \right ]^2.
\end{equation}
For the case of multi-dimensional system with quantized nuclei, the resulting mass expression in Eq. \ref{eq:gen_mass} is unchanged if we make the approximations outlined in Section \ref{sec:c_md} (i.e., that the ring-polymer position is approximated by its centroid and that contributions from multi-kink-pair configurations are neglected) and if the LZ TST is expressed in terms of the ring-polymer centroid. 

}

%

\bibliographystyle{apsrev}

%


\end{document}